\documentstyle[12pt,aps]{revtex}
\begin{document}
\title{Why is topography fractal?}
\author{Jon D. Pelletier}
\address{Department of Geological Sciences, Snee Hall, Cornell University, Ithaca, New York}
\begin{abstract}
The power spectrum $S$ of linear transects of
the earth's topography is often observed to be a power-law
function of wave number $k$ with exponent 
close to $-2$: $S(k)\propto k^{-2}$. In addition, river
networks are fractal trees that satisfy several power-law relationships
between their morphologic components. A model equation for the
evolution of the earth's topography by erosional processes which produces
fractal topography and fractal river networks 
is presented and its solutions compared in detail to real
topography. The model
is the diffusion equation for sediment transport on hillslopes and
channels with the diffusivity constant on hillslopes and proportional to
the square root of discharge in channels. The dependence of diffusivity on discharge
follows from fundamental equations of sediment transport.    
We study the model in two ways. In the first analysis the diffusivity is 
parameterized as a function of relief and a Taylor expansion procedure is
carried out to obtain a differential equation for the landform elevation
which includes the spatially-variable
diffusivity to first order in the elevation. The solution to this equation 
is a self-affine or fractal surface
with linear transects that have power spectra $S(k)\propto k^{-1.8}$, independent of the
age of the topography, consistent with observations of real topography.
The hypsometry produced by the model equation 
is skewed such that lowlands 
make up a larger fraction of the total area than highlands as observed
in real topography. 
In the second analysis we include river networks explicitly in a numerical
simulation by calculating
the discharge at every point.
We characterize the morphology of real river basins with
five independent scaling relations between six morphometric
variables. Scaling exponents are calculated for seven river networks
from a variety of tectonic environments using 
high-quality digital elevation models. 
River networks formed in our 
model match the observed scaling laws and satisfy Tokunaga side-branching statistics. 
\end{abstract}
\section{Introduction}
A remarkable feature of the earth's surface is its scale-invariance. Objects such
as a hammer or a person often need to be included in landscape photographs
because features such as variations in height of a topographic profile
have no characteristic scale. Similarly, a scale
on a map is often necessary to determine whether the map details features at the
scale of one kilometer or hundreds of kilometers.
The scale-invariance of topography can be quantified with
techniques of time series analysis. 
Variations in the height of a topographic profile
can be characterized with the probability density function
and the power spectrum.
The probability density
function quantifies how the
data is distributed about the mean. 
Two examples of probability density functions are the
normal and 
lognormal distributions. The power spectrum $S$ measures the persistence of
the data. 
The power spectrum is defined as the square of the coefficients in a Fourier
series representation of the transect. It measures the average variation
of the function at different wavelengths.
If adjacent data points are totally uncorrelated then the power spectrum
will be constant as a function of wave number (the reciprocal of the
wavelength), i.e. white noise. If adjacent values are strongly correlated relative
to points far apart the power spectrum will be large at small wave numbers (long 
wavelengths) and
small at large wave numbers (short wavelengths).
The power spectrum $S$ of linear transects
of topography have a power-law dependence on wave number with exponent close to
$-2$ over a wide range of scales: $S(k)\propto k^{-2}$ [{\it Vening Meinesz}, 1951;
{\it Mandelbrot}, 1975;
{\it Sayles and Thomas}, 1978; {\it Newman and Turcotte}, 1990].
{\it Culling and Datko} [1987] have obtained equivalent results with
the rescaled-range technique. {\it Matsushita and Ouchi} [1989] have computed the
roughness exponent $H$ defined by the relationship between the standard deviation
and the length $L$ of the transect, $\sigma\propto L^{H}$, for several
topographic transects.
They obtained $H\approx 0.55$ which implies
a power spectral exponent of $\beta=-2.1$ from the relation $\beta=2H+1$
[{\it Turcotte}, 1992]. {\it Ahnert} [1984] obtained a similar value. 
{\it Turcotte} [1987] and {\it Balmino} [1993] have 
computed the power spectrum from a spherical harmonic representation of the
earth's topography and bathymetry. They observed a power spectrum $S(k)\propto k^{-2}$
at scales less than 10,000 km and an approximately constant spectrum at
larger scales. Similar scale-invariance has been identified in the earth's bathymetry
[{\it Bell}, 1975], the topography of natural rock surfaces [{\it Brown and
Scholz}, 1985],
and the topography
of Venus [{\it Kucinskas et al.}, 1992].
The observation of scale-invariant topography on Venus indicates that
fractal topography can be formed without erosion.
A power-law power spectrum
is indicative of scale-invariance since the power-law function has no length scale
in it. {\it Turcotte} [1992] has shown that these observations define topography
to be a self-affine fractal with a fractal dimension close to 2.5.
Synthetic topography which assumes random phases in the Fourier
coefficients can be generated which have $S(k)\propto k^{-2}$ as
observed in real topography.
Images of synthetic topography with $S(k)\propto k^{-2}$ produced with techniques
described in {\it Voss} [1988] and presented in {\it Mandelbrot} [1983] resemble natural topography.
However, the lack of any river networks in these images
indicates that the power spectrum is not
a complete representation of the earth's topography [{\it Weissel, Pratson, and
Malinverno}, 1994].
{\it Gilbert} [1989] and
{\it Evans and McClean} [1995] have documented deviations from scale-invariance
of the earth's topography.

Besides the power spectrum, the distribution or hypsometry of topography
is an important statistical measure. The topography of the earth's
continents are highly skewed such that
a much larger percentage of the earth's topography is lowlands 
(topography with an elevation below the median
elevation for a region) and there
is a positive correlation between elevation and slope
(i.e. as one drives up a mountain, the steepness of the climb increases).
This is not predicted
by a model with a Gaussian distribution such as the Brownian walk 
[{\it Weissel, Pratson, and Malinverno}, 1994] .
This discrepancy between Gaussian models of topography and observed topography
is consistent with the observation of {\it Mandelbrot} [1983] who
found that when he transformed his Gaussian synthetic topography
with a cubic function, the topography looked more realistic.
 
Many studies have attempted to model the evolution of drainage networks. Recent
papers have emphasized their fractal properties.
Some authors describe discretized models which
follow the flow of discrete units
of runoff down the steepest slope and erode the hillslope with an assumed
dependence of denudation on slope (and possibly other factors)
[{\it Willgoose, Bras, and Rodriguez-Iturbe}, 1991;
{\it Chase}, 1992; {\it Kramer and Marder}, 1992;
{\it Leheny and Nagel}, 1993; {\it Inaoka and Takayasu}, 1993; {\it Howard}, 1994].
After the drainage of
a large number of randomly deposited units of precipitation, a rough landscape is
produced. {\it Willgoose et al.} [1991], {\it Leheny and Nagel} [1993],
{\it Kramer and Marder} [1992], and
{\it Inaoka and Takayasu} [1993] have shown
that their models produce drainage networks consistent with the observed scaling
behavior of real drainage networks. {\it Chase} [1992] has presented a model that can
produce fractal topography with fractal dimensions broadly consistent with observed
values. Models have also been proposed which do not explicitly model the
landform topography but only the growth of the drainage
network. {\it Stark} [1991] has proposed a model based on self-avoiding
percolation clusters.
{\it Kondoh and Matsushita} [1986], {\it Meakin, Feder, and Jossang} [1991],
{\it Masek and Turcotte} [1993], and {\it Stark} [1994]
have presented models based on diffusion-limited-aggregation (DLA) and variants
of DLA. In {\it Masek and Turcotte} [1993]
the random walkers are introduced randomly into the landscape rather than
at the boundaries (as in traditional DLA) to better model the effect of storms
producing runoff randomly in space and time.
Other approaches include that of {\it Newman and
Turcotte} [1990] who propose a cascade model similar to Kolmogorov's model of the
turbulent cascade in which the variance at each scale is dependent only on that scale
and the next largest one. Several studies have proposed continuum
growth equations. {\it Sornette and Zhang} [1993] have advocated a model equation known
as the Kardar-Parisi-Zhang (KPZ) equation, originally introduced in the physics
literature to model the growth of atomic surfaces by ion deposition
[{\it Kardar, Parisi, and Zhang}, 1986], as a model for the evolution of topography by
geomorphic processes.
{\it Newman and Turcotte} [1990] and {\it Sornette and Zhang} [1993] have
stressed the necessity of nonlinear terms in order to produce scale-invariant
topography. The KPZ equation is
\begin{equation}
\frac{\partial h}{\partial t}=D\nabla ^{2}h+\frac{\lambda}{2} (\nabla h)^{2}+
\eta (x,y,t)
\label{kpz}
\end{equation}
where $\eta (x,y,t)$ is a Gaussian white noise.
The first term represents the classic Culling diffusion model of slope erosion.
{\it Culling} [1960, 1963]
hypothesized that the horizontal flux of eroded material
was proportional to the slope. With conservation of mass this yields the
diffusion equation. Solutions to the diffusion equation
have been successfully applied to modeling the evolution of alluvial fans,
incised channels, prograding deltas, and eroding fault scarps [{\it Wallace}, 1977;
{\it Nash}, 1980;
{\it Begin et al.}, 1981; {\it Gill}, 1983a,b; {\it Hanks et al.}, 1984;
{\it Hanks and Wallace}, 1985; {\it Kenyon
and Turcotte}, 1985; {\it Phillips and Sutherland}, 1986]. {\it McKean et al.}
[1993] have implied diffusion coefficients
of hillslope evolution with geochemical methods. They obtained similar diffusion
constants to those
inferred from modeling fault scarp relaxation.
The last term in equation \ref{kpz} represents spatial variations in erosion intensity produced by
the intermittent nature of runoff production and mass movements, spatial
variations in the erodibility of the soil and rock of the landscape, and
episodic tectonic uplift.
The inclusion of a stochastic term such as this is a universal feature of models
considered in the physics literature which produce self-affine surfaces.
{\it Sornette and Zhang} [1993] have argued that the nonlinear term in equation
\ref{kpz} is essential for the
formation of a scale-invariant surface.
The nonlinear term is equivalent to assuming that the erosion
rate is proportional to the local exposed landscape surface. Somfai and Sander
[Scaling and river networks: A Landau theory, unpublished manuscript, 1996]
have also employed the KPZ equation and have produced
landscapes which obey Horton's laws with simulations
incorporating only the nonlinear term in the the equation for the local erosion rate
of the surface. {\it Smith and Bretherton} [1972] have presented a generalized
model of hillslope evolution which includes the KPZ nonlinear term as a
special case. 
{\it Giacometti, Maritan, and Banavar} [1995] have
proposed a continuum growth equation that includes higher order terms not
present in the KPZ equation.

It should be emphasized that a great variety of
models, including most of those listed, obey Horton's laws. Horton's laws state
that stream number, average stream length, and average drainage area decrease
geometrically with stream order defined by the Strahler ordering scheme.
Since so many different models obey Horton's laws,
agreement with Horton's laws does not
appear to be sufficient to verify that a model is an accurate
representation of drainage network formation [{\it Willgoose}, 1994].
In fact, {\it Kirchner} [1993] has argued that Horton's laws are satisfied
by virtually all possible branching networks. {\it Tokunaga} [1984] and
{\it Peckham's} [1995]
improved classification of stream order appears to be a more stringent test
for models and is a significant advance in this regard.
 
In this paper we investigate a model of landform evolution with topographic evolution
by geomorphic processes 
parameterized with the diffusion equation. The model we propose reproduces the
observed statistical behavior of both topography and river networks.
The diffusion equation models the relaxation of topographic variations by overland and
channel flow. In Section 2 we 
present a derivation due to {\it Begin, Meyer, and Schumm} [1981] which
shows diffusivity to be proportional to the square root of the discharge.  
It is this dependence on discharge, itself dependent on the basin morphology,
that gives rise to a complex unstable behavior from such a simple model. 
Our first analysis of
the model will be to parameterize the diffusivity as a function of relief 
and perform a Taylor expansion retaining the spatial variability of diffusivity to
first order in the elevation. A nonlinear partial differential equation is derived and
solved which reproduces the hypsometry and power-spectral behavior
of real topography. In Section 4 numerical simulations are performed which
explicitly include river networks and calculate the discharge at every point.
Realistic river networks and topography are generated. 

\section{Dependence of Diffusivity on Discharge}  
{\it Begin, Meyer, and Schumm} [1981] have derived the diffusion equation for 
channel
evolution using similar assumptions to those used by {\it Culling} [1960, 1963]
for hillslope evolution.
Their derivation began with
classic equations of sediment transport. They showed diffusivity
to be proportional to the discharge per unit width above a point on the surface.
The authors began with a commonly observed
empirical relationship between sediment flux $Q_{s}$ and bottom shear stress $\tau$:
\begin{equation}
Q_{s}\propto\tau^{\frac{3}{2}}
\label{weak}
\end{equation}
The mean flow velocity above a point on the surface is
\begin{equation}
v=\Big(\frac{8}{f}gR\frac{\partial h}{\partial x}\Big)^{\frac{1}{2}}
\label{lame}
\end{equation}
where $f$ is the Darcy-Weisbach factor, $g$ is the acceleration due to gravity, $R$ is the
hydraulic radius, and $\frac{\partial h}{\partial x}$ is the channel slope along the length
of the longitudinal profile of a channel.
For wide channels, $v=Q_{w}/R$ where $Q_{w}$ is the water discharge per unit width.
Substituting this relation into equation \ref{lame}, {\it Begin, Meyer, and Schumm} [1981] obtained
\begin{equation}
R=Q_{w}^{2}\Big(\frac{f}{8g\frac{\partial h}{\partial x}}\Big)^{\frac{1}{3}}
\end{equation}
Substituting the expression for $R$ into the equation $\tau=\rho gR\frac{\partial h}{\partial x}$,
where $\rho$ is the density of water, gives
\begin{equation}
\tau\propto\Big(\frac{\partial h}{\partial x}\Big)^{\frac{2}{3}}
\end{equation}
Substituting this into equation \ref{weak} results in the following expression for sediment discharge
\begin{equation}
Q_{s}\propto Q_{w}\frac{\partial h}{\partial x}
\label{done}
\end{equation}
Since $\frac{\partial h}{\partial t}= \frac{\partial Q_{s}}{\partial x}$ by conservation of mass,
equation \ref{done} gives a diffusion equation for the channel elevation with a diffusivity
propotional to the water discharge per unit width:
\begin{equation}
\frac{\partial h}{\partial t}= cQ_{w}\frac{\partial^{2}h}{\partial x^{2}}
\end{equation}
where $c$ is a constant dependent on $f$, $g$, and $\rho$.
The diffusion model and a discharge-dependent diffusivity is consistent with the laboratory
measurements of aggradation and degradation in channels by {\it Gill} [1983a,b] and {\it Phillips
and Sutherland} [1986].
 
Since the width of a river is proportional to the square root
of the discharge [{\it Leopold, Wolman, and Miller}, 1964], the diffusivity of a
channel
is proportional to $Q/w\propto Q/Q^{\frac{1}{2}}\propto Q^{\frac{1}{2}}$, 
the square root of the discharge.
In our model we will  
have a constant diffusivity on hillslopes and a diffusivity proportional to
the square root of discharge in channels:
\begin{eqnarray}
\frac{\partial h}{\partial t}=\nabla(D(Q)\nabla h) \nonumber\\
D=\mbox{if the site is a hillslope}\nonumber\\
D\propto Q^{\frac{1}{2}}\ \ \mbox{if the site is a channel}
\end{eqnarray}

In order to solve this equation it is necessary to parameterize discharge
in terms of the morphology of the basin.
Discharge is principally a function of drainage basin area. It is often 
assumed that discharge and area are proportional [{\it Rodriguez-Iturbe et al.},
1992]. However, in an analysis of 350 of the world's largest river basins,
{\it Mulder and Syvitski} [1996] have established the
relationship between average discharge and drainage area 
to be a power-law relationship with exponent
$0.75$: $Q_{av}\propto A^{0.75}$. This means that larger basins drain
less water per unit area than small basins. One interpretation of this observation
is that more infiltration occurs in large basins.
To test this hypothesis, {\it Mulder and Syvitski} [1996]
related the average discharge $Q_{av}$ to the basin area corrected by the
cosine of the average basin slope, $A/\cos\alpha$. They obtained a correlation 
coefficient of 0.9 with this expression compared to 0.74 for the power-law
relationship, lending support to the hypothesis. 
However, in order to maintain consistency with the power-law
relationships we will identify between the morphometric variables, we will
relate discharge to drainage area using the power-law relationship
$Q_{av}\propto A^{\frac{3}{4}}$. 
The model equation as a function of area is then
\begin{eqnarray}
\frac{\partial h}{\partial t}=\nabla(D(A)\nabla h) \nonumber\\
D=D_{h}\ \ \mbox{if the site is a hillslope}\nonumber\\
D\propto (Q^{\frac{1}{2}})^{\frac{3}{4}} =D_{c}A^{\frac{3}{8}}\ \ \mbox{if the site is a channel}
\label{kpztransformed}
\end{eqnarray}
 
\section{Approximate Solution of the Model Equation} 
Discharge is a function of basin relief, the difference in elevation between the
highest and lowest points in the basin. 
In first-order streams high in mountainous areas the discharge is
very small compared to that for lowland rivers. In this section we parameterize
the discharge as a function of relief to obtain a single differential 
equation for the local elevation in space and time, $h(x,y,t)$. 
The equation is solved and its 
solutions are found to have a hypsometry and power spectrum comparable to 
those of real topography dominated by erosional processes.

In Section 4 we will present the results of morphometric analyses
of river networks. We will show the drainage area
to be a power function with exponents between 3 and 5 
of the basin relief, $h-h_{max}$, where
$h$ is the elevation of the outlet and
$h_{max}$ is the maximum elevation of the basin. Since the diffusivity is
defined to be a power-law function of drainage area, the dependence of
the
diffusivity on relief is also a power law with exponent $a$: 
\begin{equation}
D\propto(h-h_{max})^{a}
\label{approxi}
\end{equation} 
$D$ can be expanded in a Taylor series for small $h$: $D\approx D_{0}-D_{1}h$
where $D_{0}$ and $D_{1}$ are positive constants. 

The diffusion equation with spatially variable diffusivity is 
\begin{eqnarray} 
\frac{\partial h}{\partial t}=\nabla(D(h/h_{max})\nabla h) \nonumber\\
=D(h/h_{max})\nabla^{2} h+\frac{\partial D}{\partial h}(\nabla h)^{2} 
\label{newkpz}
\end{eqnarray} 
The diffusivity must be kept inside the gradient term since it is
not a constant. The chain rule has been used.

Substituting equation \ref{approxi} into equation \ref{newkpz} and keeping only
terms first order in $h$ gives
\begin{equation}   
\frac{\partial h}{\partial t}=D_{0}\nabla^{2} h-D_{1}(\nabla h)^{2}
\label{suck}
\end{equation}

The effects of the terms in equation \ref{suck} are illustrated in
Figures 1 and 2. In Figure 1a, the topography at time $t$
is assumed to
be given by
a Gaussian function. The topography is taken to be one-dimensional for the
purposes of illustration. The rate of change of the surface
resulting from the diffusion term is given in Figure 1b and the
surface at time $t+\Delta t$ is shown in Figure 1c. The effect of the
diffusion term is to aggrade the topography where the surface is concave
up and erode where the surface is concave down. The topography of the 
surface at time $t+\Delta t$ is also a Gaussian function.

In contrast, the nonlinear term does not preserve the Gaussianity of the initial
surface. Figure 2 presents the same sequence of graphs for
the nonlinear term. The term erodes material where the slope is large
resulting in a more concave topography with a skewed hypsometry and a
larger fraction of topography
in lowlands compared to the original Gaussian function.   

Equation \ref{suck} is completely deterministic. However,
there is abundant empirical evidence that spatially and temporally
variable erosion rates are universal over a wide range of time and length scales.
Much of the sediment carried away in rivers is carried away in intermittent
storms whose occurrence can only be described statistically. An example of this
intermittency is the time series of sediment load in the Santa Clara River. The
sediment load of the Santa Clara was carefully monitored for a period of 18
years. Over half of the total sediment yield carried by the river was transported
in only three large floods totaling seven days [{\it Milliman and Syvitski}, 1992].
Similar large bursts of sediment transport are evident in the sediment transport time
series of {\it Plotnick and Prestegaard} [1993] on time scales of hours.
Many examples of large floods which have resulted in major landscape modification have
been documented in the geologic record [{\it Bretz}, 1969;
{\it Meyer and Nash}, 1983; {\it Ager} 1993].
Spatially and temporally variable erosion rates have been documented over a
variety of spatial and temporal scales
and geomorphological settings [{\it Luk}, 1982; {\it Ormi}, 1982;
{\it Schmidt}, 1985].
A steady-state competition between intermittent stochastic forcing
and subsequent relaxation of the landscape
has been argued to be the essential dynamic of landscape evolution
[{\it Wolman and Gerson}, 1978]. A model of landscape evolution as a sequence of
epsodic 
events is consistent with the age distributions of rock avalanches and sedimentary
sequences [{\it Griffiths}, 1993].
These observations suggest that any model of landform evolution must include stochastic 
spatial and 
temporal variability in erosion rates. If we include spatial and temporal variations
in erosion rates by adding a Gaussian white noise erosion rate, $\eta (x,y,t)$, to equation \ref{suck},
the result is the KPZ equation (equation \ref{kpz}). 

The solution to the KPZ equation is a
self-affine or fractal surface with linear transects that have power-law
power spectra
with an exponent of $-1.8$:
$S(k)\propto k^{-1.8}$ [{\it Amar and Family}, 1989].
This is concluded by relating the Hausdorff measure
reported for the KPZ model, $H\approx 0.4$, to the one-dimensional power
spectral exponent, $\beta$, through the relation $\beta=2H+1$ [{\it Turcotte}, 1992].
Due to the difficulty of solving nonlinear partial differential equations numerically
[Newman and Bray, Strong-coupling behavior in discrete Kardar-Parisi-Zhang equations,
unpublished manuscript, 1996], a cellular automaton model,
the restricted solid-on-solid (RSOS) model, has been developed by {\it Kim and Kosterlitz} [1989]
with rules that mimic the terms in the KPZ equation. {\it Park and Kahng} [1995] have shown that
the RSOS model is equivalent to the KPZ equation in the continuum limit.
In the RSOS model, a site on a
two-dimensional lattice of
points is chosen at random. The height of the surface at that point is incremented by one
if the elevation at that point is greater than or equal to the elevations of all of its four
nearest neighbors.
If this restriction is not satisfied, nothing happens.
This rule is repeated until a surface with a height equal
to or greater than the linear dimension of the lattice is generated. Periodic boundary
conditions are used. We performed simulations of this model on a 256 x 256 lattice.
A shaded relief image of an example of
a surface generated with this model is shown in Figure 3.
The average power spectrum, estimated as the
square of the 
coefficients of the Fast Fourier Transform, of linear transects of
this surface is presented in Figure 4. 
The power spectrum of each row of the
lattice was computed and then averaged with the power spectra of all other rows in order
to obtain the average power spectrum. A good match with the power spectrum $S(k)\propto k^{-1.8}$,
indicated by the straight line, is obtained. This power spectrum agrees with the finite difference
calculation of the KPZ equation of {\it Amar and Family} [1989].
This power spectrum is close to the spectrum $S(k)\propto k^{-2}$ observed in
erosional topography by {\it Huang and Turcotte} [1989] and others. The
power spectral behavior observed in real topography 
is independent of the age of the topography (time since significant uplift occurred) and the
initial relief following tectonic uplift.
For instance, young, rough mountain ranges, such as the Rocky Mountains,
exhibit the same power spectral exponent or fractal dimension
as smooth mountain ranges such as the Appalachians.
Similarly, in the RSOS model a steady-state condition is acheived once 
a rough surface is produced. In the steady-state condition, the smoothing effects of the diffusion
term are balanced by the roughening effects and
the power spectral behavior is independent of time. 
A rougher landscape, defined as a larger variance per unit wavelength,
can be produced by increasing the ratio of the variance
of the stochastic term to the diffusion constant.
Rougher topography has a larger power spectral density. However, the power spectral
exponent, which quantifies the relative amplitude of topography at different
wavelengths, is the same for rough or smooth topography.

The topography of the earth's surface
is skewed such that
a much larger percentage of the earth's topography is lowlands and there
is a positive correlation between elevation and slope.
This skew can be associated with the nonlinear term in equation 11. 
If only the diffusion term were present, the resulting topography would
have a Gaussian distribution. This is because any linear transformation
of a function with a Gaussian distribution, such as the noise term in
equation 1, results in a function with a Gaussian distribution. 
The probability density function (p.d.f.) of elevations (hypsometry)
produced by the RSOS model
is presented in
Figure 5a. 
The p.d.f. was computed using the surfaces generated from 
10 simulation runs. The observed p.d.f. is not Gaussian, but is skewed
slightly such that the most probable elevation is below the mean elevation
of the landscape, i.e. more of the total landscape area is represented by
lowlands than highlands. 
The skew of the distribution increases as the ratio $D_{1}/D_{0}$ increases.
The probability distributions for the Kentucky
and Mississippi River basins are given in Figures 5b and 5c, respectively. 
The ETOPO5 dataset [{\it Loughridge}, 1986] was used to compute the 
hypsometry of the Mississippi River basin and the USGS 1$\deg$ DEMs 
[{\it United States Geological Survey}, 1990] were
used to compute the hypsometry of the Kentucky River basin.  
Both hypsometries exhibit skew towards lower elevations. This skew
is directly comparable to the skew in the RSOS topography and can be
associated with the nonlinear term.

Figure
6a-c shows the
cumulative percentage of area larger than a given area computed by {\it Harrison et al.} [1983]
using the ETOPO5 dataset
[{\it Loughridge}, 1986] for the continents of Africa, North America, and South
America, with
a least-square fit to a lognormal distribution. The hypsometric curves presented in
the independent study of {\it Cogley}
[1985] are similar to those presented in {\it Harrison et al.} [1983].
The calculated hypsometric curves appear to approach a skewed lognormal distribution as the age since
significant uplift
increases. Of the three continents, Africa has experienced the
least neotectonic uplift. South America has experienced the most. The hyposmetry of
Africa matches a lognormal distribution more closely than South America where the presence of
the relatively young Andes mountains results in
a significant deviation 
from a lognormal distribution. This suggests that continental hypsometric curves approach
a
lognormal distribution as erosion has more time to act on the landscape.

\section{River Networks}
The objective of this section is to incorporate river networks into 
equation \ref{kpztransformed}.
There are several
power-law relationships between the morphologic components of
river basins. Some of these components are based on the Strahler ordering scheme.
In this scheme, a stream with a channel head is defined as a first-order stream.
When two like-order streams combine they form a downstream segment one order higher
than the order of the tributary streams.
A partial list of the observed morphological relations in natural river networks is:
 
1) {\it Horton} [1945] defined three ratios, $R_{B}$, $R_{L}$, and $R_{A}$ to be the ratio
between the number, average length, and average drainage area from the streams of one order to
those of the next lowest order. He found this ratio to be a constant for all stream
orders. The fractal dimension of river networks is defined as $D=\log R_{B}/\log R_{L}$.
$D$ is usually found to be approximately 1.9 [{\it Turcotte}, 1992].
 
2) An improved classification scheme leading to a relation similar to Horton's laws has been developed by
{\it Tokunaga} [1984] and {\it Peckham} [1995]. They defined matrix elements $T_{o,k}$
as the number of side tributaries of order $k$ of streams of order $o$.
Natural river networks satisfy the constraint that $T_{o,o-k}$ is constant
for all $o$. Shreve's classic random topology model [{\it Shreve}, 1966] fails to satisfy this
constraint [{\it Peckham}, 1995]. The classic DLA growth model satisfies this property [{\it Ossadnik}, 1992].
 
3) {\it Hack} [1957] found that the length of a main channel length scales with the
drainage area according to power law: $L\propto A^{q}$. He reported values of $q\approx 0.6$
for two river basins. 
{\it Gray} [1961] obtained a value of $0.57$.
Some other morphometric analyses based on hundreds of river basins, however, have found no significant
deviation from $0.5$ [{\it Montgomery and Dietrich}, 1992; {\it Mulder and
Syvitski}, 1996].
 
4) Along-channel slope is a power law function of discharge
with exponent close to $-1/2$: $S\propto Q_{av}^{-\frac{1}{2}}$ [{\it Carlston},
1968]. If it
is assumed that discharge and area are proportional, this implies that 
channel slope is inversely proportional to the square root of the drainage 
area: $S\propto A^{-\frac{1}{2}}$. This assumption is often made and    
the relationship $S\propto A^{-\frac{1}{2}}$ is considered to be 
a universal feature of river networks [{\it Tarboton et al.}, 1989]. 
However, as we have pointed out,
average discharge and drainage area are not proportional. Thus, one of the  
two relationships $S\propto Q_{av}^{-\frac{1}{2}}$ or  
$S\propto A^{-\frac{1}{2}}$ should be considered suspect.
 
The fractal properties of river networks have reviewed by many authors
including {\it LaBarbera and Rosso} [1989], {\it Beer and Borgas} [1993], {\it Nikora} [1994],
and {\it Maritan et al.} [1996].
{\it Abrahams} [1984] has reviewed other emprical relations observed for river basin
morphology such as the statistics of junction angles.
 
In order to better characterize the morphometric relations between drainage 
basin components we have carried out river network extraction and analyses 
on seven basins from a variety of tectonic environments using high-quality 
Digital Elevation Models. Four river networks were chosen from the
composite DEM of {\it Fielding et al.} [1994]. This
data set has 80 m resolution and does not rely on the interpolation
of contour lines. Such interpolation, as is done in the USGS 1$\deg$ DEMs,
may lead to biased slope estimates.  
Three of these basins are located along the Himalayan front in Nepal,
Kumaun, and Bhutan. The fourth is located in the Shanxi Province, China
and is formed in loess. This basin has an unusally ordered shape characterized
by a high degree of symmetry and unusually straight valleys. The remaining three
networks are located in North America. The Kentucky River basin and
Schoharie Creek basin were extracted from USGS 1$\deg$ DEMs [{\it 
United States Geological Survey}, 1990]. The Mississippi River basin was
chosen so that a large basin was represented. The
Mississippi was extracted from the ETOPO5 data set [{\it Loughridge}, 1986].
The seven basins are plotted in Figure 7. Although the river network extraction
and analyses were carried out down to the pixel size of the DEM, only rivers 
with Strahler orders larger than three
were plotted so that the network can be identified. 
The river network extraction and analyses were carried out with RiverTools 1.01
[{\it Peckham}, 1997]. 

The analyses we carried out enable us to identify five independent 
morphologic relationships between six components. The results are shown
in Figures 8 through 12 and are summarized in Table 1. The six
morphometric components are stream number of a particular Strahler order
$N$, Strahler order $o$, main channel length $L$, drainage area $A$,
basin relief $R$, and along-channel slope $S$. The relationships between the
variables are defined as
\begin{eqnarray}
N\propto A^{p} \\
L\propto A^{q} \\
S\propto A^{r} \\ 
R\propto A^{s} \\
A\propto t^{o}
\end{eqnarray}

Figure 8 presents the total number of streams of a given Strahler order
as a function of the average basin area for that Strahler order for each
of the seven basins. The order of plots in Figures 8-13 is, from top to bottom, the
Kumaun basin, the Loess plateau of the Shanxi Province, Schoharie Creek,
the Nepal basin, the Kentucky River basin, the Mississippi River basin, and the 
Bhutan river basin. The plots are offset so that they may be placed on the
same graph. Figure 8 indicates that $N$ is approximately proportional to
$A^{-1}$ indicating that $p\approx -1$. 

Figure 9 presents the relationship between $L$ and $A$. The plots indicate
that $q\approx 0.5$. There has been considerable debate about whether 
$q$ approaches 0.5 exactly or whether there
is a significant deviation. The reason for the debate is that if $q\neq 1/2$
then this may represent a deviation from self-similarity [{\it Ijjasz-Vasquez et al.},
1993]. Our analyses exhibit a variation in $q$ from basin to basin that is 
roughly equal to the previously reported devations 
from $1/2$. Therefore, we cannot 
definitively conclude whether $q$ differs from $1/2$ in river basins
as a general rule.

Figure 10 indicates that $r\approx 3/8$. This is inconsistent with previous
studies that have reported $S\propto A^{-\frac{1}{2}}$ [{\it Tarboton et al.}, 1989]. 
The value $r\approx 3/8$
is entirely consistent with the observed relationships of channel slope to
discharge and the scaling of discharge and drainage area: $S\propto 
Q_{av}^{-\frac{1}{2}}$ and $Q_{av}\propto A^{\frac{3}{4}}$ implies
$S\propto A^{\frac{3}{8}}$, as observed. 
The work of {\it Tarboton et al.} [1989] carried out analyses on two
river basins with three orders-of-magnitude of area in the analyses.   
Given that our analyses were carried out on several basins 
with five orders-of-magnitude of area with a high-quality DEM  
we propose that $r\approx 3/8$ is a more reliable estimate. 
This conclusion appears to be consistent with the data of {\it Montgomery
and Dietrich} [1988] who presented similar slope-area relationships. Although
no exponents were obtained, the trends of their data follow a power-law
relationship with exponent significantly greater than $-1/2$ (i.e. closer
to $-3/8$). 

The observed scaling between along-channel slope and drainage area
follows directly from equation 9. The flux of sediment is given by
$Q_{s}=D_{c}A^{\frac{3}{8}}\frac{\partial h}{\partial x}$.
For a longitudinal profile in equilibrium,
$\frac{\partial Q_{s}}{\partial x}=0$.
$\frac{\partial Q_{s}}{\partial x}=0$ implies that
the average
slope $\frac{\partial h}{\partial x}$
must be related to drainage area $A$ as $\frac{\partial h}{\partial x}
\propto A^{-\frac{3}{8}}$ as observed.  

Plotted in Figure 11 are the relationships between relief $R$ and area $A$.
A range of values is observed between $s\approx 1/5$ and $s\approx 1/3$.

The relationship between basin area and Strahler order $o$ is plotted
in Figure 12. In this figure area is plotted on a logarithmic (base 10)
scale while Strahler order is plotted on a linear scale. The slopes
are observed to be approximately $2/3$. Thus, the Horton ratio $R_{A}$ 
is equal to 10$^{\frac{2}{3}}\approx$ 4.6. Using the other morphometric
relations, we can estimate $R_{L}$ and $R_{B}$ as 2.2 and 4.6, respectively.      

The results of Tokunaga side-branching statistical analyses on the seven
drainage basins is presented in Figure 13. In this ordering scheme matrix
elements $T_{o,k}$ are defined to be 
the number of side tributaries of Strahler order $k$ of streams of Strahler
order $o$.
Natural river networks satisfy the constraint that $T_{o,o-k}$ is constant
for all $o$. We have determined $T_{k}$ by averaging the vales of 
$T_{o,o-k}$ over $o$:
\begin{equation}
T_{k}=\frac{1}{n-k}\sum_{o=1}^{n-k}T_{o,o-k}
\end{equation}
Plotted in Figure 13 is $T_{k}$ on a logarithmic (base 10) scale
as a function of $k$ on a linear scale. The plots indicate that 
$T_{k}\propto u^{k}$ where $u$ is estimated to be $10^{0.4}\approx 2.5$. 
  
We now consider the numerical simulation of equation 9. The 
resulting model river networks will be analyzed in the same manner as the
real river networks analyzed above. 
The equation was 
simulated on a square lattice of 64 x 64 grid points. The equation was
discretized in space and time. Integration in time utilized the 
predictor-corrector method which varies the time step to ensure stability.
The initial condition was an elevation of 1.0 on every point of 
lattice except for the upper left corner which was moved down to 0.0 at
$t$=0 and fixed to be zero for the entire calculation. The slope   
at all of the edges was fixed to be 0.1 except for the upper left corner which
was unconstrained. This constraint on slope at the boundaries is necessary
because either the elevation or its derivative must be specified at
the boundaries. Since the largest slopes in a basin are in mountainous
streams far from the basin outlet, imposing a significant slope on the boundary
gridpoints while allowing the elevations to be unconstrained as the drainage
divide advances or retreats appeared to be the most realistic   
boundary condition.  
The
contributing area at each point was continuously updated. Each grid point
drained to the lowest of its nearest neighbors. If a grid point was moved
up or down such that it no longer was the lowest neighboring grid point
for one of its neighbors, the contributing area of
each grid point downhill of the old drainage 
path was decremented by one and the contributing area of each grid point
downhill of the new drainage path was incremented by one.   

Since we are modeling the hillslope and the river network as distinct states, we must model the growth
of the channels into the hillslope. It may be of interest to model fluctuations in the topography
with a river network already formed, but a complete model of landscape evolution
must consider the feedback between a growing
network and the topography of the adjacent hillslopes.
One simple way to model growth of the network is to start a simulation with one or more
small channels draining to the
border of the lattice and extend the channel from these locations at a given rate
when the area drained by the grid point, a proxy
for discharge, exceeds a given threshold. Such a model of headward growth is supported by
field studies [{\it Patton and Schumm}, 1975;
{\it Begin and Schumm}, 1979]. It was also used in the model of
{\it Willgoose, Bras, and Rodriguez-Iturbe} [1991].
In our model the threshold for channelization was set to be zero so that
every point in the lattice eventually became a channel. The rate
of channel advancement is governed by conservation of mass. For the
channel to advance a distance $\Delta x$, the mass that must be moved
downstream is equal to $\frac{\partial h}{\partial x}\Delta x$ where
$\frac{\partial h}{\partial x}$ is the slope along the longitudinal
profile. Since the flux is equal to $D(A)\frac{\partial h}{\partial x}$,
the rate of channel advancement is proportional to $D(A)$.

Greyscale plots of the elevations of the model surface for four
instants of time are plotted in Figure 14a-d. The surface elevations
are mapped to a brightness scale with a gamme function with a coefficient of
2.0. The four instants of time correspond to those where (a) 1/8 , (b) 1/4, (c)   
1/2, and (d) all of the lattice has become channelized. Despite the fact 
that the simulation is fully deterministic and begins with symmetric 
boundary conditions, an asymmetric basin morphology is produced as a result
of the instability of channel downcutting. As a river basin cuts down its valley,
the river basin increases its drainage area. This further enhances channel
downcutting and so one. Thus, any difference in basin drainage area tends to
be amplified over time by this instability.

In Figure 15a, the river network corresponding to the surface 
of Figure 14d is presented. In this plot the river width is made equal
to the Strahler order so that thicker rivers indicate those that drain
more area. In Figure 15b is plotted the river network created with the same
parameters as Figure 15a but with some stochastic variability included in 
the sediment transport. For each pixel and each time step, a factor
$(1+0.1\eta)$ was multiplied times $\Delta h$, where $\eta$ is a Gaussian
white noise with mean zero and standard deviation of one. As previously
argued, spatial and temporal variations in erodibility are a universal
feature of landscape evolution. The purpose of including this stochastic 
variability was to determine its effect, if any, on the morphology of the
basin.

Figures 16-21 are plots of the morphologic relationships corresponding to 
those for real river basins presented in Figure 8-13. In these plots three
river basins are analyzed. The results for the deterministic simulation of
resulting in the basin of Figure 14d are presented as the top graph. The
middle graph represents the results of morphometric analyses on the
partially-developed river basin of Figure 14c. The results of the stochastic
river basin illustrated in Figure 15b are presented as the bottom graph.
The results of the morphmetric analyses on the real and model river basins 
are summarized in Table 1.

The same scaling relations are observed for the three model river basins, 
suggesting that the scaling laws we have identified are satisfied 
by river basins continuously as they evolve by headward growth and that
they are satisfied even in the presence of heterogeneous erodibility and/or
runoff. Close agreement between model and real river basins was observed for
most of the morphometric relationships. The scaling exponent of slope $S$
and area $A$ was observed to be slightly higher for the model river networks
(-0.29 to -0.32) compared to the observed value of approximately $-0.37$ in real
river basins.
This was unexpected since the result $S\propto A^{-0.37}$ is satisfied 
exactly for a river basin in equilibrium (where the sediment flux is everywhere
constant). The reason for this discrepancy will require further study. 
In addition, the Horton ratio $R_{A}$ and the Tokunaga exponent $u$ quantified
in Figures 20 and 21 were about 20\% higher for the model basins than for the
real river basins. Nevertheless, the broad consistency between the observed 
relations and those satisfied by the model suggests that this model captures
the essential dynamics of self-organization in river basins. 

The probability density function of the model basin of Figure 15a is
presented in Figure 22. As observed in the Kentucky and Mississippi  
River basins, the hypsometry is skewed such that the most probable elevation
is lower than the median elevation. The evolution of the hypsometric curve
(cumulative distribution function) for the model basins of Figure 14a-d is
shown in Figure 23 as the top to bottom graphs, respectively. A close
similarity exists between this sequence and the evolution observed in
real basins. As a example, in Figure 24 we show the hypsometric curves 
observed over time at Perth Amboy, New Jersey by {\it Schumm} [1956]. Both the
real and model 
basins begin with a nearly constant surface equal to the maximum elevation
of the basin. As the channels dissect the basin, the hypsometric curve 
smooths out. The basins reach equilibrium with a skewed p.d.f. as evidenced
by the fact that the model and real hypsometric curves fall below $1/2$ at
a normalized elevation of $1/2$. A Gaussian or any other symmetric distribution
would pass through ($1/2$,$1/2$).       

\section{Conclusions}
We have presented a model of landscape evolution by overland and channel
flow which reproduces many of the basic statistical features of topography
and river basin morphology. The model is based on the observation that
channel and hillslope adjustment by sediment transport can often be
modeled by the diffusion equation with a diffusivity dependent on discharge.
The dependence of diffusivity on discharge introduces a nonlinear term in
the partial differential equation for landscape elevation which is responsible
for the fractal nature of topographic transects and results in a characteristic
hypsometric curve. The morphometric relationships for seven river basins have
been computed and have been found to be remarkably universal. Model drainage
basins have morphologies which closely approximate those observed in nature.   

\section{Acknowledgements}
I wish to thank Chris Duncan and Don Turcotte
for helpful conversations. 
Special thanks is extended to Scott Peckham for making RiverTools 1.01,
his remarkable 
river network and extraction program, available over the Internet.
Most of this work would not have been possible without Scott's program.
This work was supported in part by
NASA grant NAGW-4702.

\begin{figure}
\caption{Illustration of the topographic evolution with diffusion. (a) The initial
topographic profile at some time $t$, 
given by a Gaussian, (b) the Laplacian of the initial
topography, and (c) the initial topography (dashed line) and final topography
(solid line) after a small increment of time according to the diffusion equation.
The topography aggrades where the surface was concave and is eroded where the
surface was concave down.     
After {\it Barabasi and Stanley} [1995].}  
\end{figure}

\begin{figure}
\caption{Illustration of the topographic evolution with the nonlinear term
of equation 11. (a) The initial topographic profile at some time $t$, given
by a Gaussian, (b) the nonlinear term of the initial topography, and (c)
the initial topography (dashed line) and final topography
(solid line) after a small increment of time. The final surface is skewed
so that it is no longer Gaussian and has more area in lowlands than the
inital profile.  
After {\it Barabasi and Stanley} [1995].}  
\end{figure}

\begin{figure}
\caption{Shaded relief image of topography produced with the 
restricted solid-on-solid (RSOS) model.}
\end{figure} 

\begin{figure}
\caption{Average power spectrum $S$ as a function of wave number $k$ for
one-dimensional transects of the surface generated with the RSOS model.
A least-square fit to the logarithms of the ordinate and abscissa yield
a slope of $-1.81$ indicating that $S(k)\propto k^{-1.81}$.}
\end{figure} 

\begin{figure}
\caption{(a) Probability density function of elevations of the surface (hypsometry)
produced by the RSOS model.
(b) Probability density function of the Kentucky River basin. (c) Probability density function of the 
Mississippi River basin.
All three exhibit a significant skew such that lowlands (topography below the
median elevation) make up a larger fraction of the total area than highlands.} 
\end{figure}

\begin{figure}
\caption{Cumulative hypsometric curves (dots), the fraction of area above an elevation normalized
by the maximum
elevation, for (a) Africa, (b) North America, and (c) South America, from the data of {\it Harrison et al.}
[1983]. The line accompanying each plot is the least squares fit to a lognormal distribution for each
continent.}
\label{hypsometry}
\end{figure}

\begin{figure}
\caption{Drainage networks analyzed: (a) Kumaun, (b) Loess Plateau, Shanxi 
Province, (c) Schoharie Creek, (d) Nepal, (e) Kentucky River, (f) 
Mississippi River, (g) Bhutan}  
\end{figure}

\begin{figure}
\caption{Plot of average number of streams $N$ of a given Strahler order as 
a function of the average drainage area $A$ for that Strahler order for each of
the seven river basins. The plots are offset from one another so that they
may be placed on the same graph. The plots correspond, from top to bottom,
to the river basins in (a) through (g) of Figure 7, respectively.
The data indicate that $N\approx A^{-1}$.} 
\end{figure}

\begin{figure}   
\caption{Plot of average main channel length $L$
of a given Strahler order as 
a function of the average drainage area $A$ for that Strahler order for each of  
the seven river basins. The plots are offset from one another so that they  
may be placed on the same graph. The plots correspond, from top to bottom, 
to the river basins in (a) through (g) of Figure 7, respectively. 
The data indicate that $L\approx A^{q}$ with $q\approx$ 0.5-0.6.} 
\end{figure}

\begin{figure}   
\caption{Plot of average along-channel slope $S$ of a given Strahler order as 
a function of the average drainage area $A$ for that Strahler order for each of  
the seven river basins. The plots are offset from one another so that they  
may be placed on the same graph. The plots correspond, from top to bottom, 
to the river basins in (a) through (g) of Figure 7, respectively. 
The data indicate that $S\approx A^{-\frac{3}{8}}$.} 
\end{figure}

\begin{figure}   
\caption{Plot of average basin relief $R$ for a given Strahler order as 
a function of the average drainage area $A$ for that Strahler order for each of  
the seven river basins. The plots are offset from one another so that they  
may be placed on the same graph. The plots correspond, from top to bottom, 
to the river basins in (a) through (g) of Figure 7, respectively. 
The data indicate that $R\approx A^{-s}$ with $s$ ranging from 1/5 to 1/3.} 
\end{figure}

\begin{figure}   
\caption{Plot of average basin area $A$ of a given Strahler order as 
a function of the Strahler order $o$ for each of  
the seven river basins. Note the log-linear scale.
The plots are offset from one another so that they  
may be placed on the same graph. 
The data indicate that $R_{A}$ is constant and equal to approximately
10$^{\frac{2}{3}}\approx$ 4.6.} 
\end{figure}

\begin{figure}   
\caption{Plot of the Tokunaga ratio $T_{k}$ 
a function of the Strahler order $k$ for each of
the seven river basins. Note the log-linear scale.
The plots are offset from one another so that they
may be placed on the same graph. 
The data indicate that 
$T_{k}\propto u^{k}$ with $u\approx$ 10$^{0.4} \approx$ 2.5.}
\end{figure}

\begin{figure}
\caption{Greyscale plot of the elevation of the model river network 
when (a) 1/8, (b) 1/4, (c) 1/2, and (d) all of the grid points of the 
64 x 64 lattice has become channelized. The elevations are mapped to 
brightness scale with a gamma fucntion with a coefficient of 2.0.}
\end{figure}

\begin{figure} 
\caption{Model river networks produced after all of the grid points have
become channels. The width of the river is proportional to its order.
The fully deterministic model is shown in (a). A model run where the
diffusivity is allowed to have a stochastic variation with a standard
deviation 10\% of the mean is shown in (b).}
\end{figure} 

\begin{figure}
\caption{Plot of average number of streams $N$ of a given Strahler order as
a function of the average drainage area $A$ for that Strahler order for 
three model river basins. The plots are offset from one another so that they
may be placed on the same graph. The plots correspond, from top to bottom,
to the model river basins produced with a deterministic 64 x 64 model run
until all of the grid points were channels, a deterministic 64 x 64 model run
until 50\% of the grid points were channels, and
a 64 x 64 model with small (10\%) stochastic varaitions in the diffusivity
run until all of the grid points were channels. 
The data indicate that $N\approx A^{-1}$, similar
to that observed for real river networks.}
\end{figure}

\begin{figure}
\caption{Plot of average main channel length $L$
of a given Strahler order as
a function of the average drainage area $A$ for that Strahler order for each of
the three model river basins.
The plots are in the same order as Figure 16.
The data indicate that $L\approx A^{q}$ with $q\approx$ 0.5=0.6, similar
to that observed for real river networks.}
\end{figure}
\clearpage 
\begin{figure}
\caption{Plot of average along-channel slope $S$ of a given Strahler order as
a function of the average drainage area $A$ for that Strahler order for each of
the three model river basins. 
The plots are in the same order as Figure 16.
The data indicate that $S\approx A^{-0.3}$, a slightly larger exponent than
that observed in real river basins.}
\end{figure}
 
\begin{figure}
\caption{Plot of average basin relief $R$ for a given Strahler order as
a function of the average drainage area $A$ for that Strahler order for each of
the three model river basins. 
The data indicate that $R\approx A^{-s}$ with $s\approx$0.3, consistent with
observations.}
\end{figure}
 
\begin{figure}
\caption{Plot of average basin area $A$ of a given Strahler order as
a function of the Strahler order $o$ for each of
the three model river basins. Note the log-linear scale.
The data indicate that $R_{A}$ is constant and equal to approximately
10$^{0.75}\approx$ 5.6, slightly larger than the observed value of 4.6.}
\end{figure}

\begin{figure}
\caption{Plot of the Tokunaga ratio $T_{k}$
a function of the Strahler order $k$ for each of
the three model river basins. Note the log-linear scale.
The data indicate that
$T_{k}\propto u^{k}$ with $u\approx$ 10$^{0.5} \approx$ 3.2.}
\end{figure}

\begin{figure}
\caption{Probability density function of the model river basin
of Figure 17a. The skew in this p.d.f. is directly analagous to
that of the Kentucky and Mississippi River basins of Figure 5b and
5c, respectively, and can be associated with the dependence of
diffusivity on elevation.}  
\end{figure}

\begin{figure} 
\caption{Evolution of the hypsometric curves for the
model river basin for the four instants of time 
illustrated in Figure 14. Increasing time results in a
smoother hypsometric curve.} 
\end{figure} 

\begin{figure} 
\caption{Observed evolution of the hypsometric curves from 
field observations at Perth Amboy, New Jersey [{\it Schumm}, 1956].} 
\end{figure}

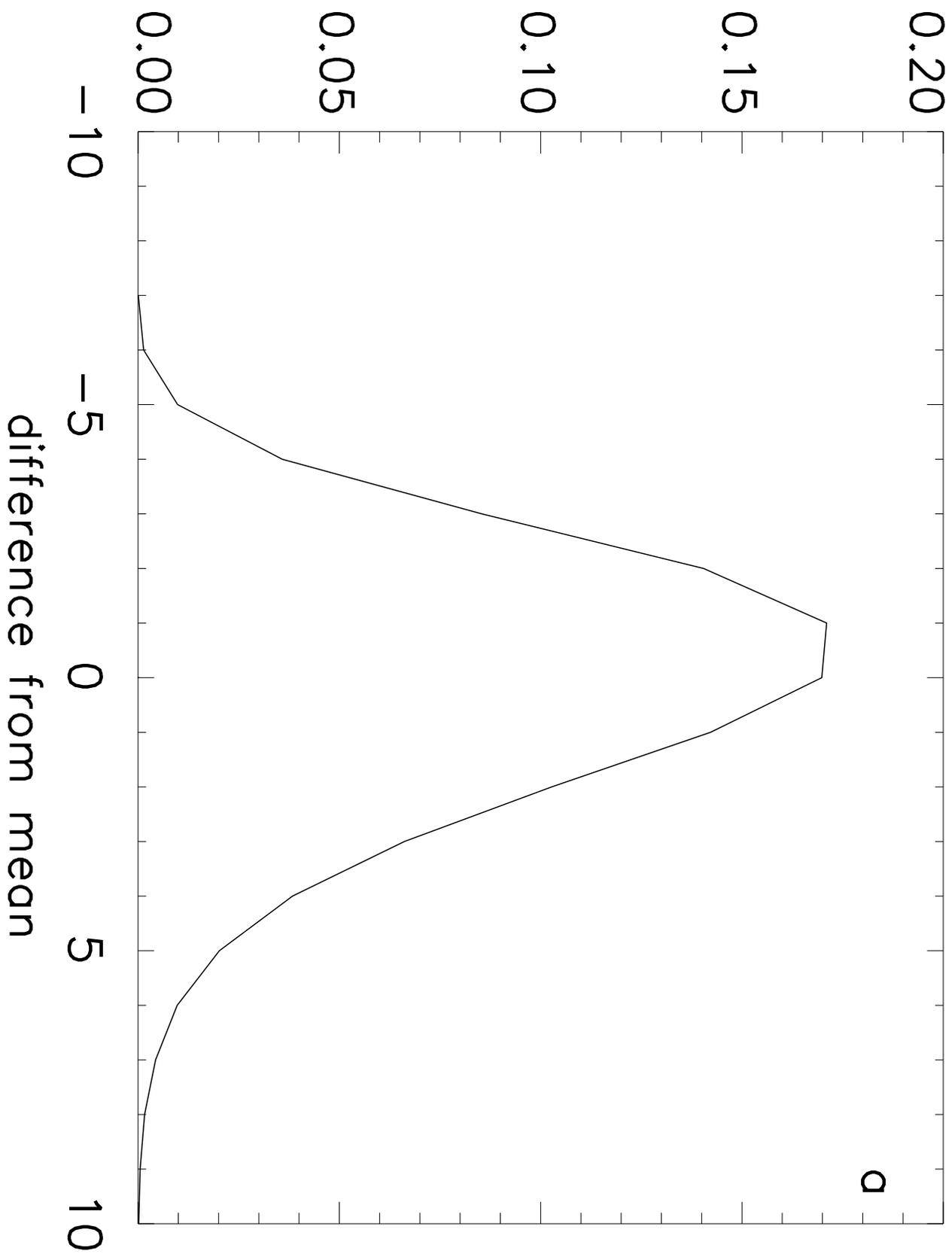
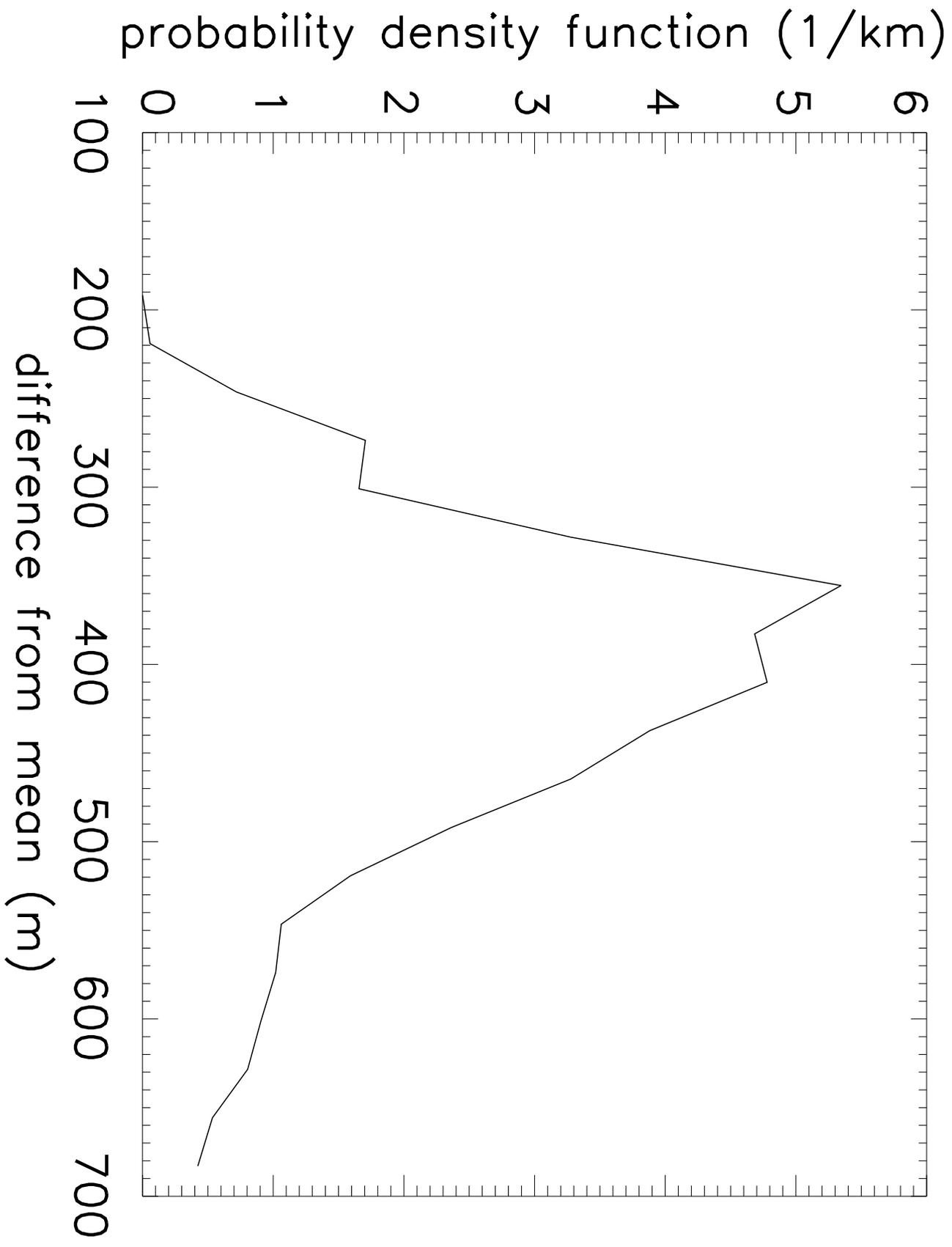
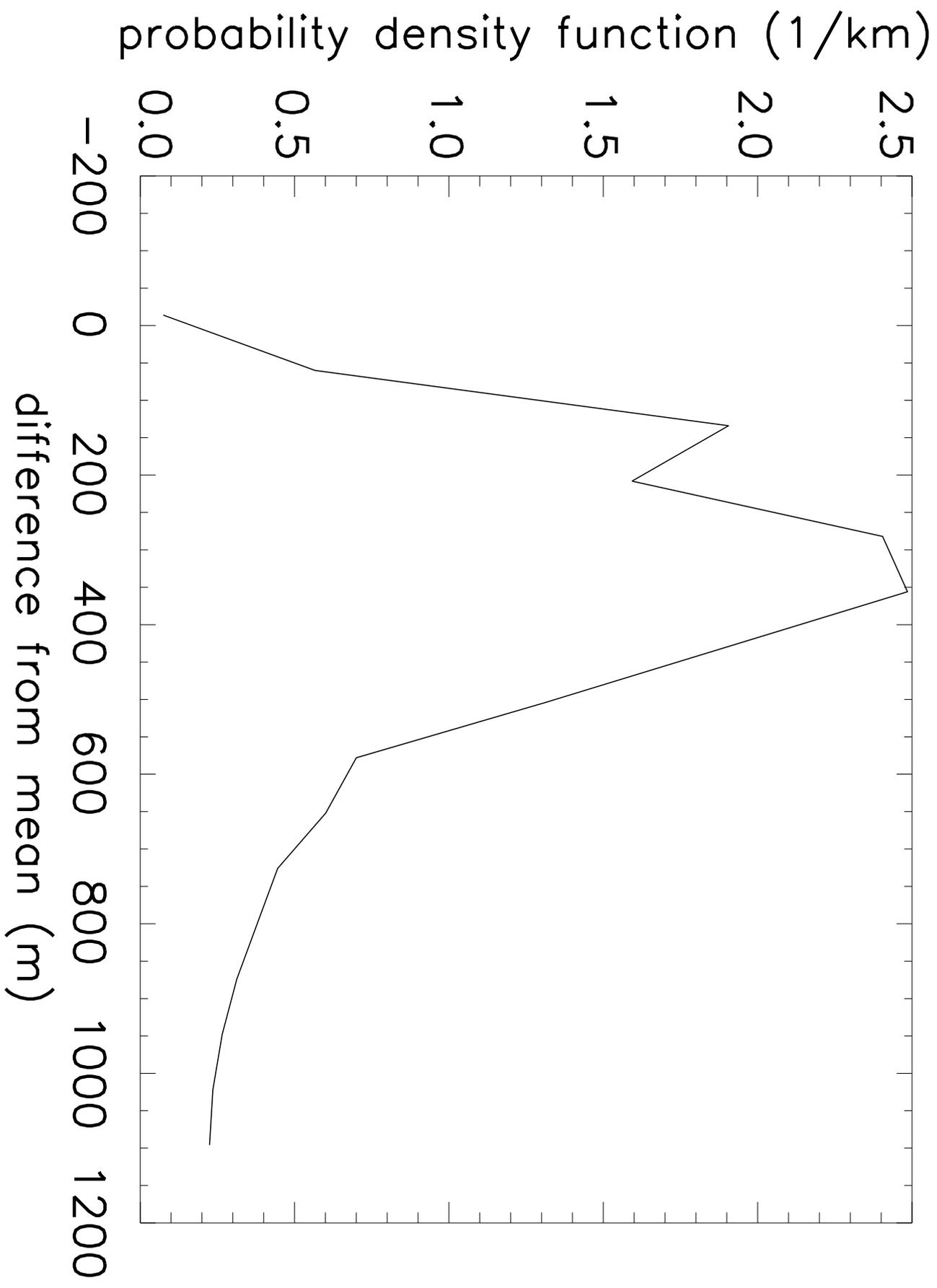
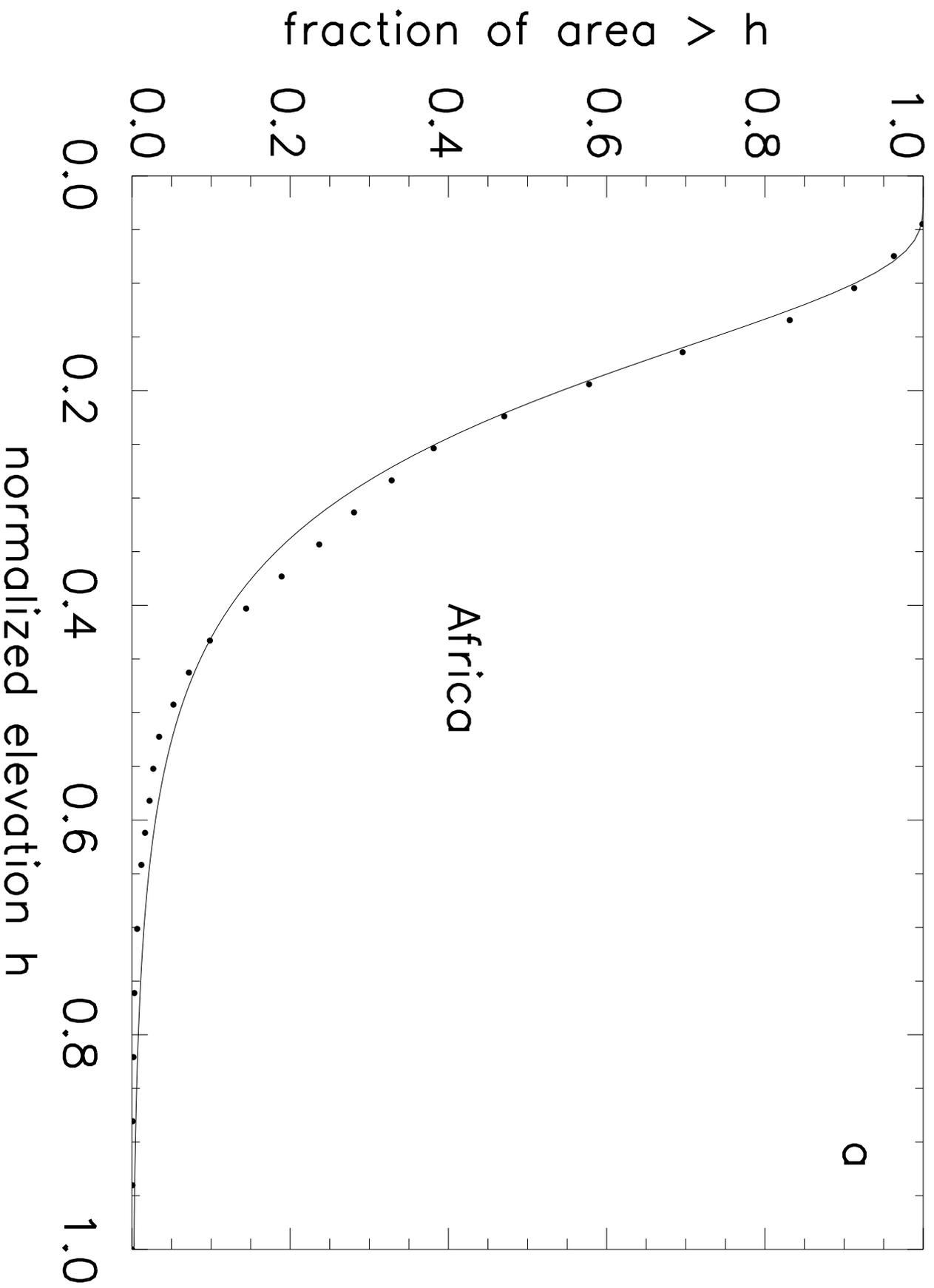
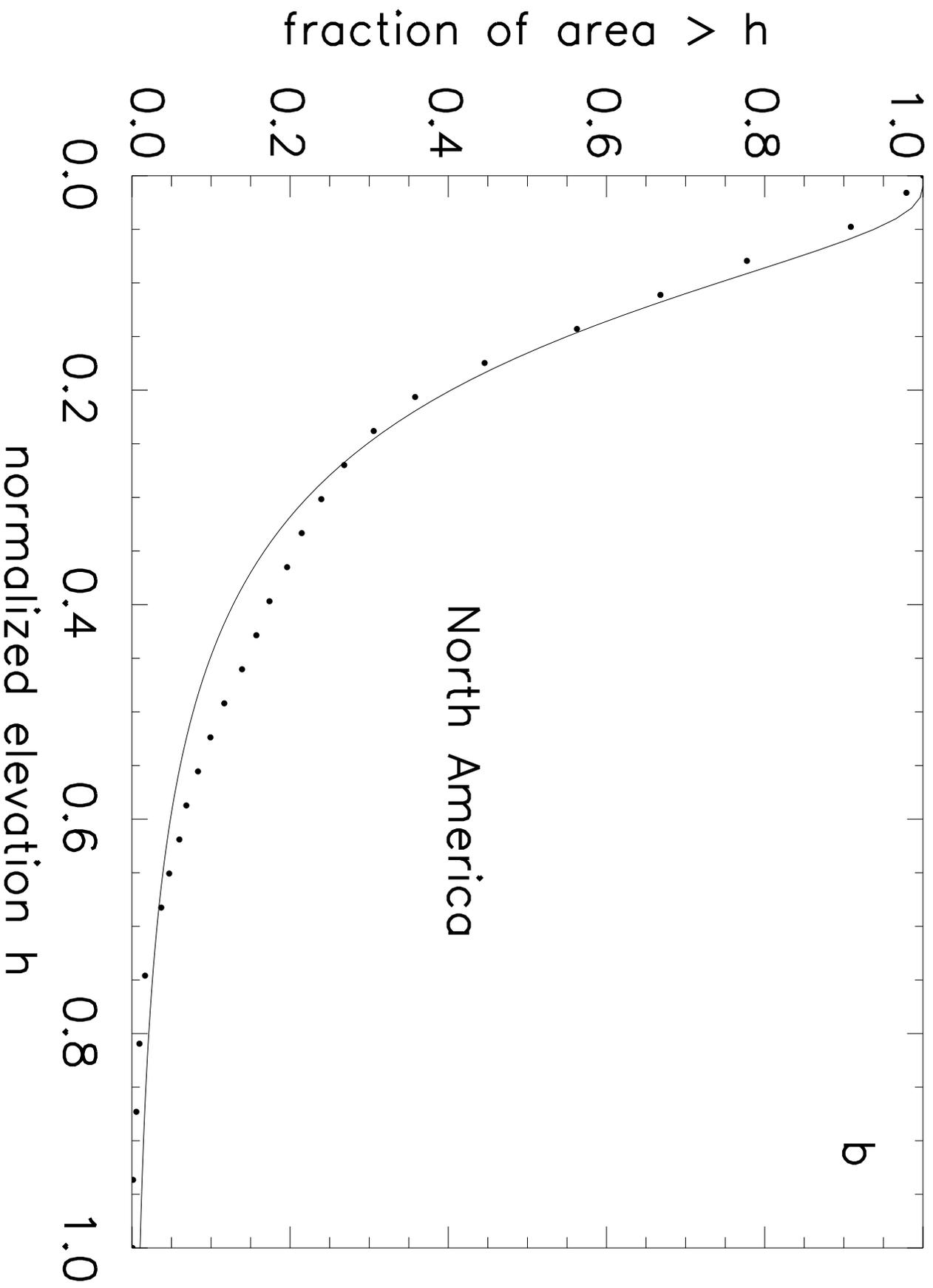
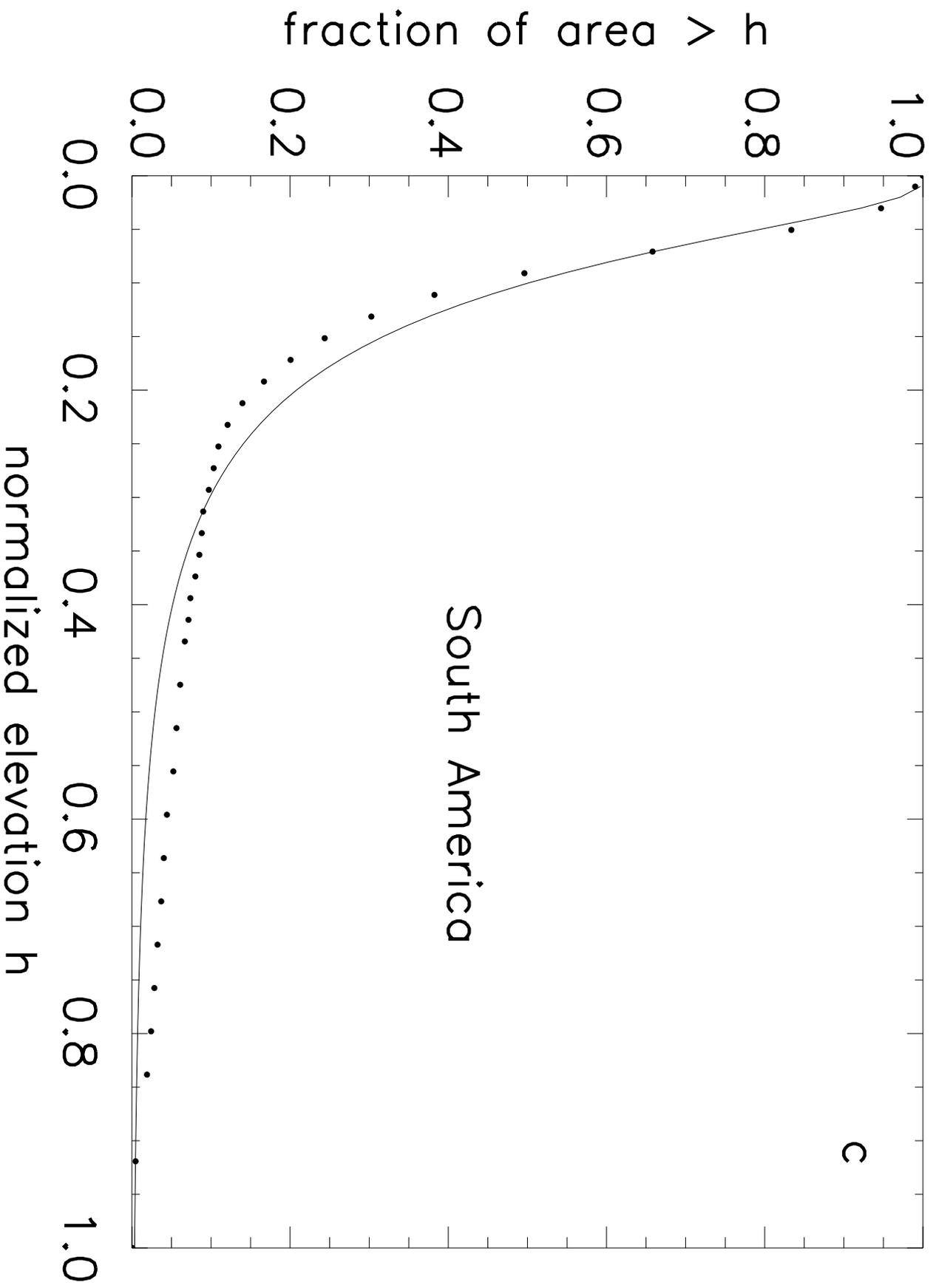
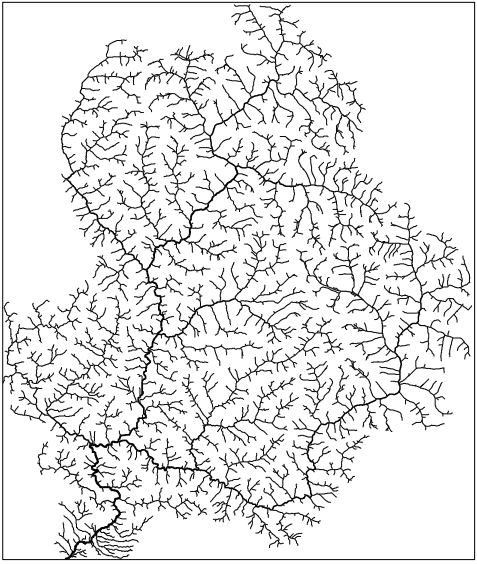
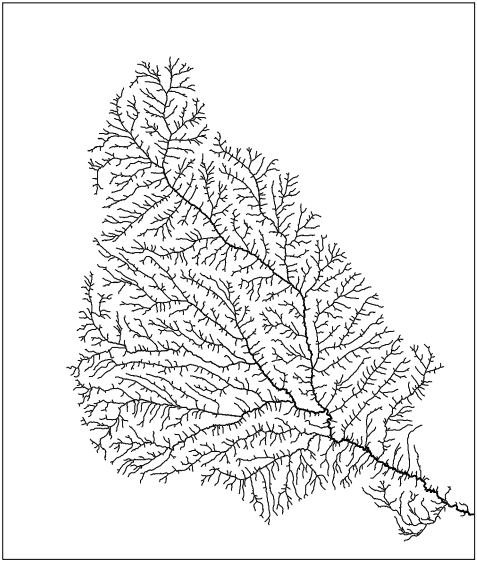
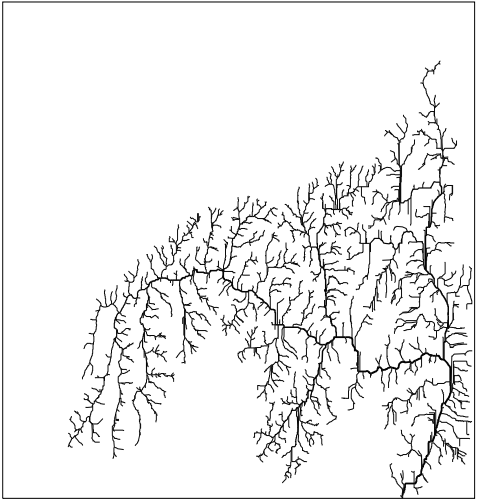
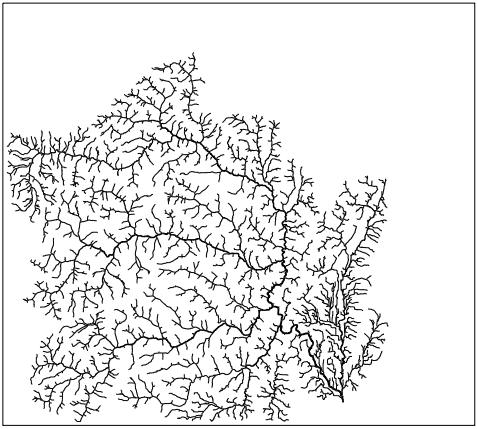
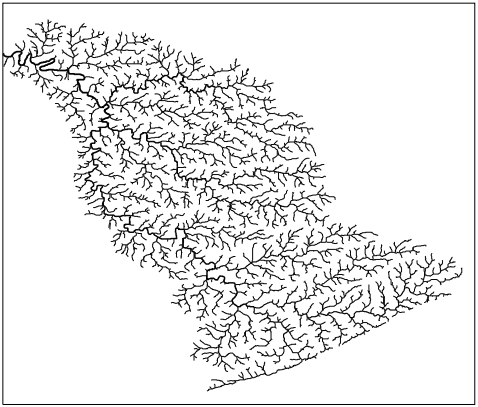
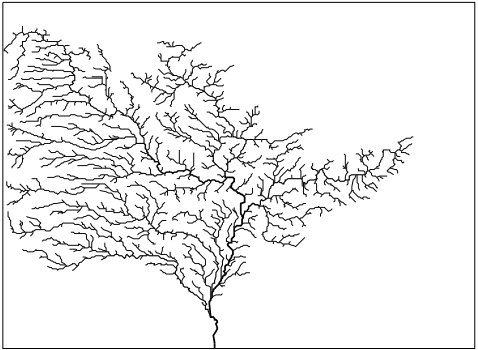
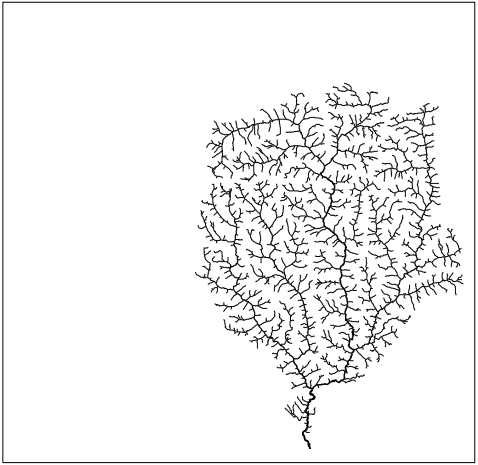
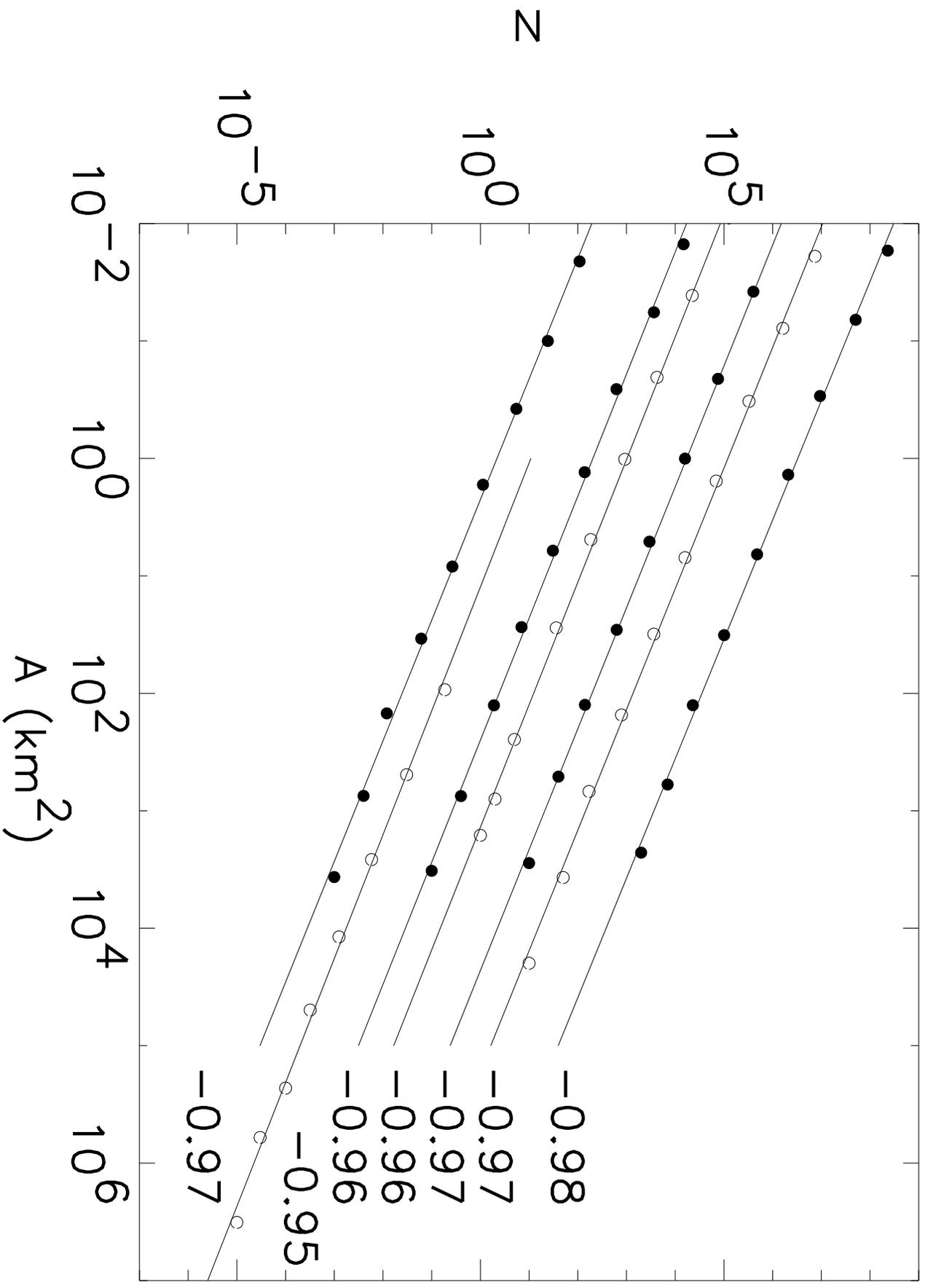
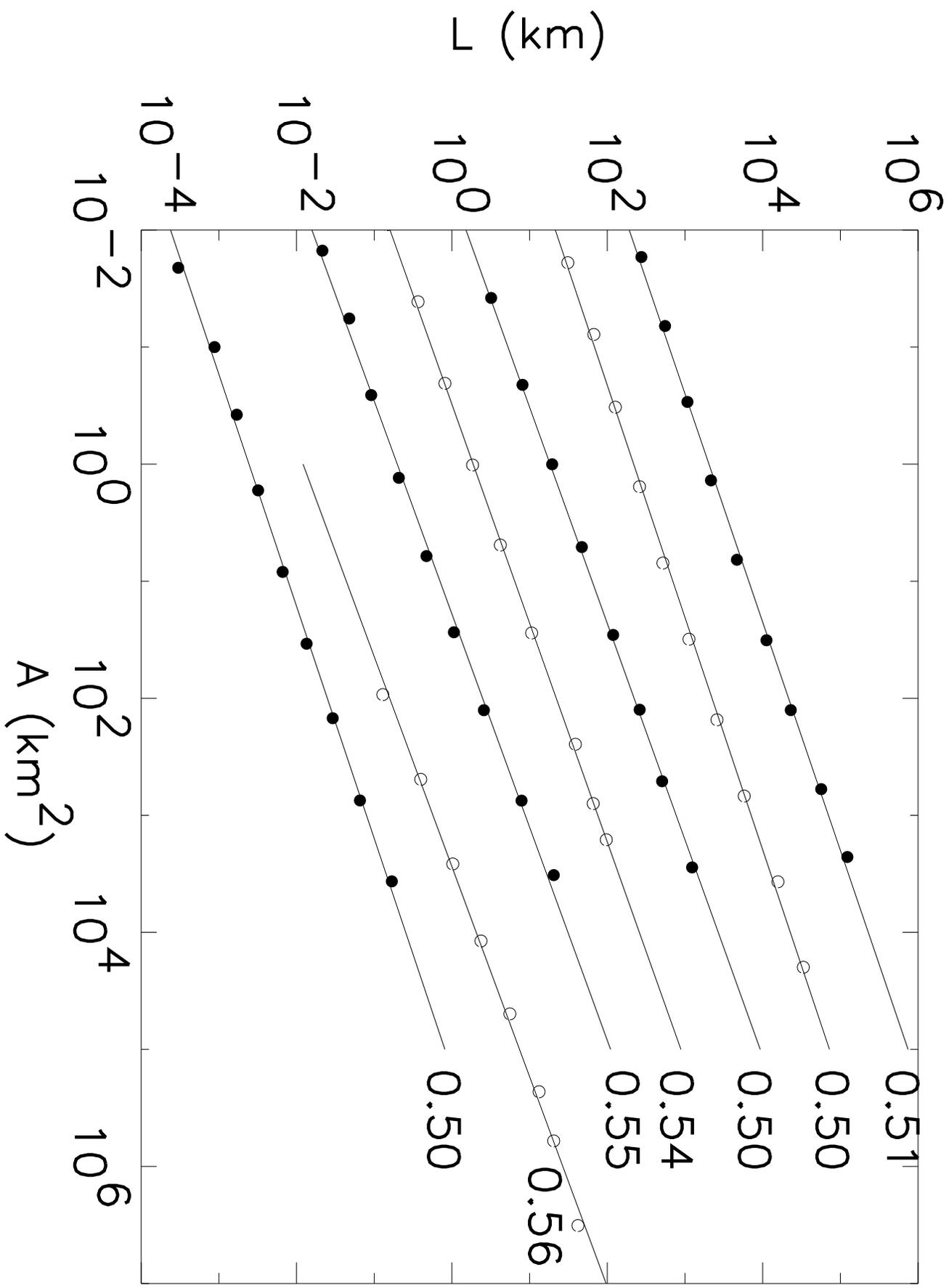
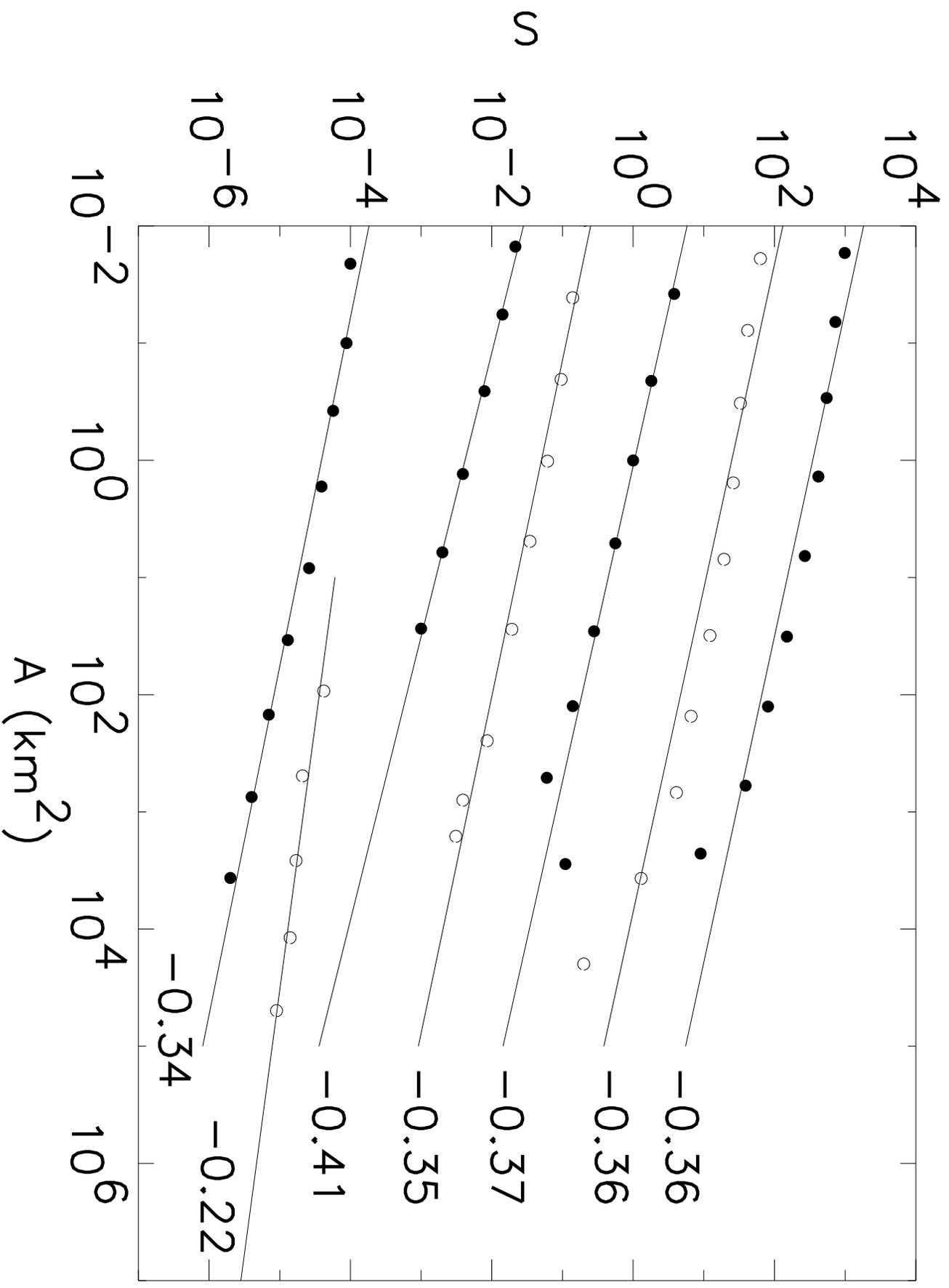
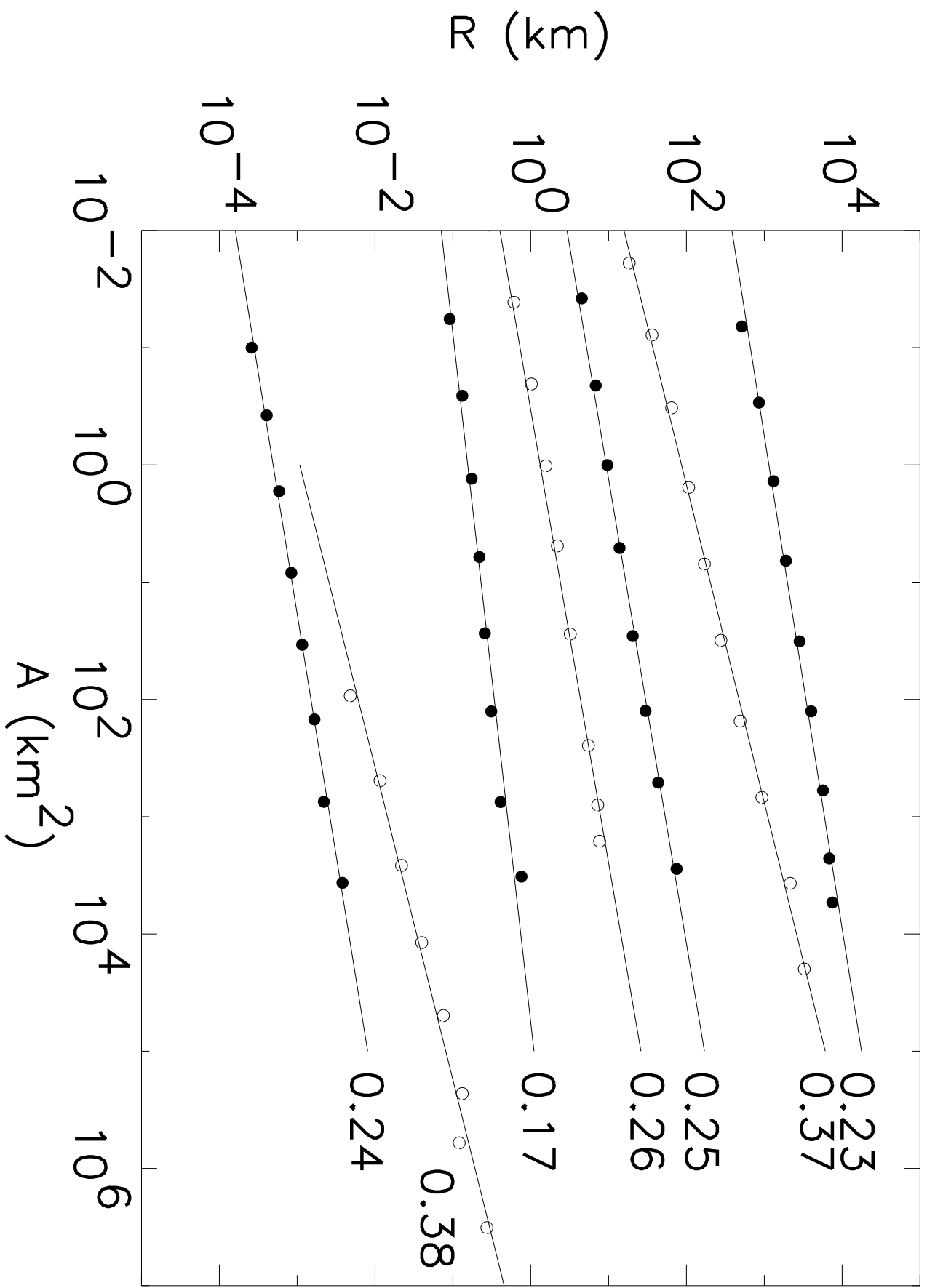
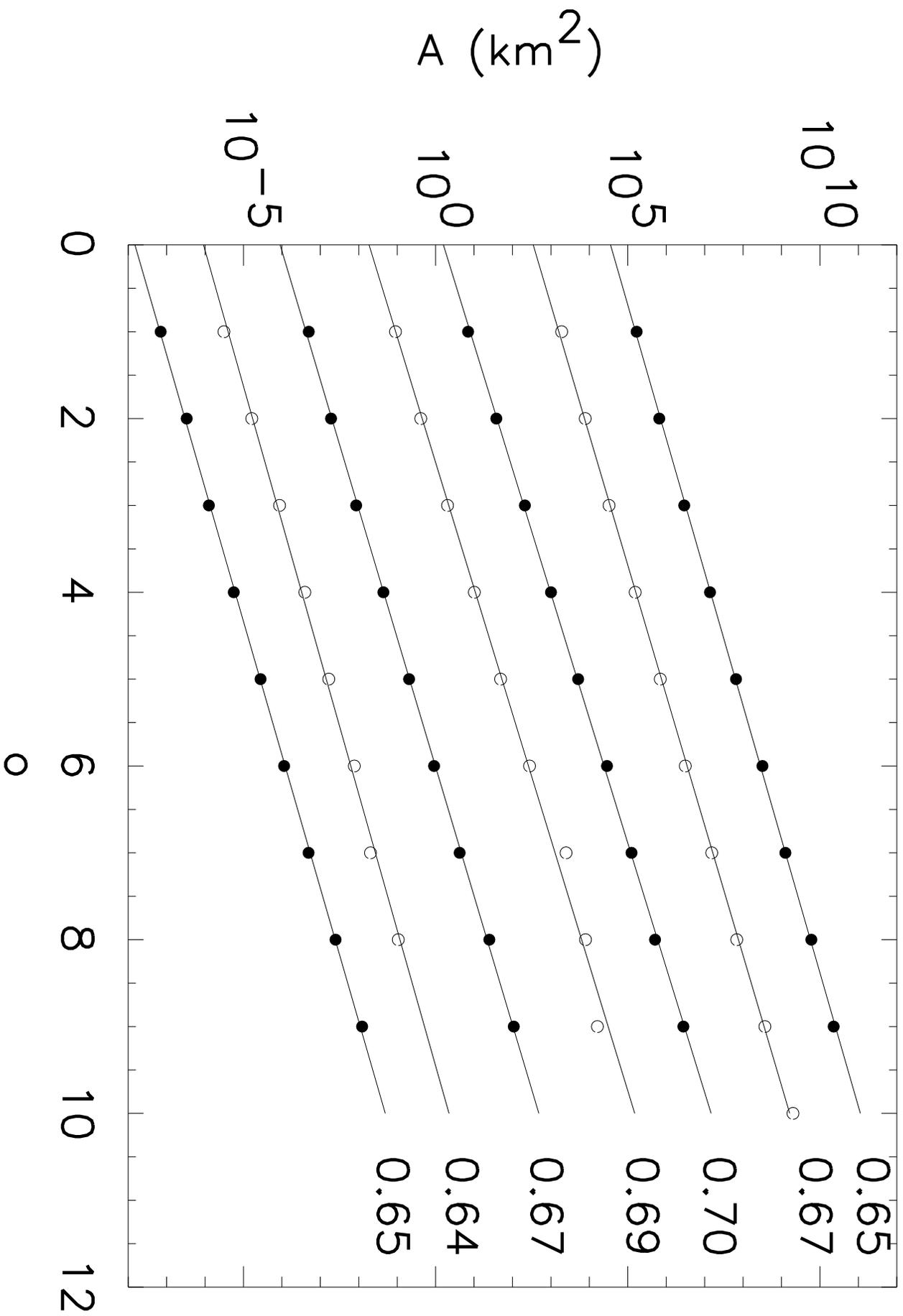
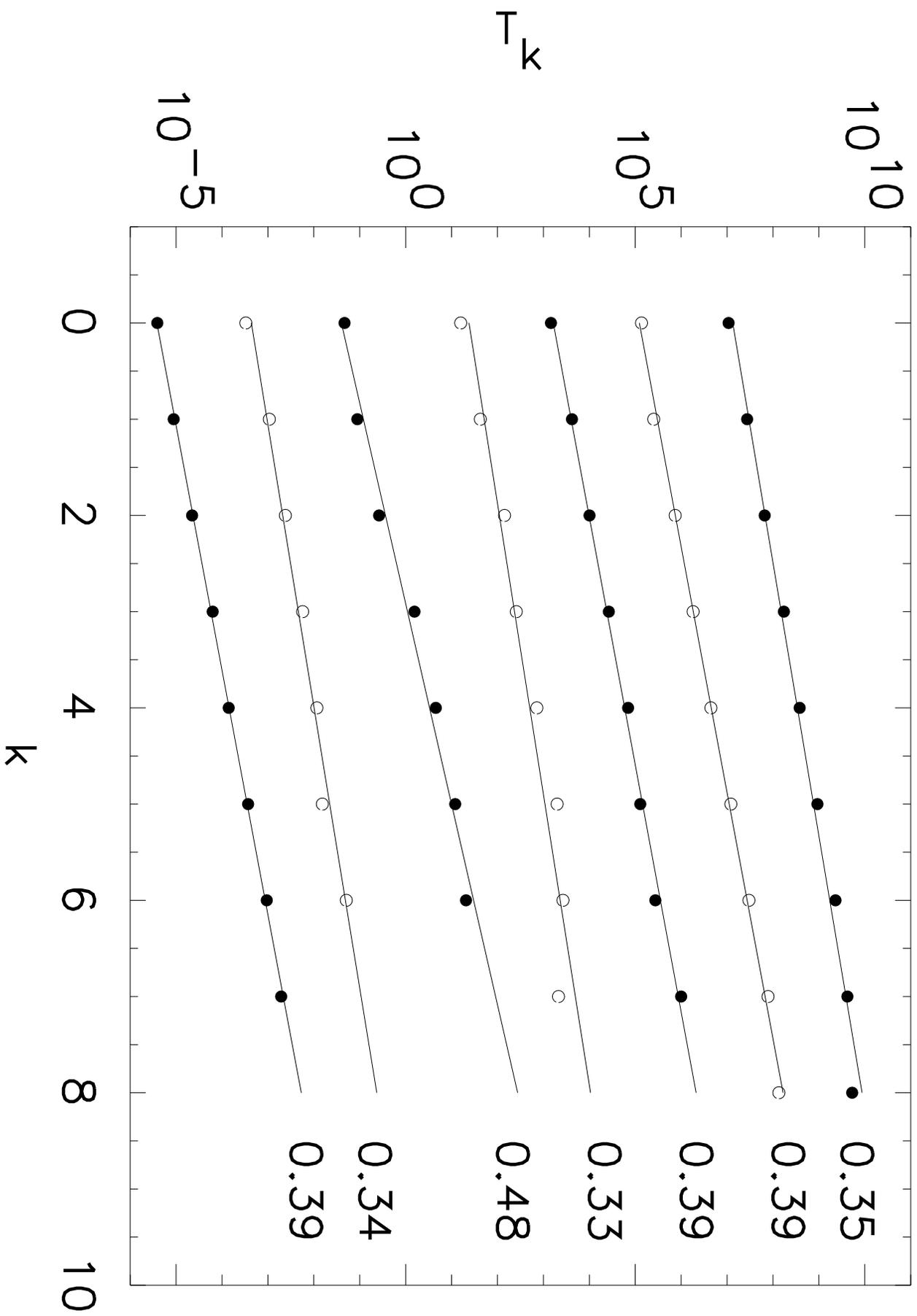
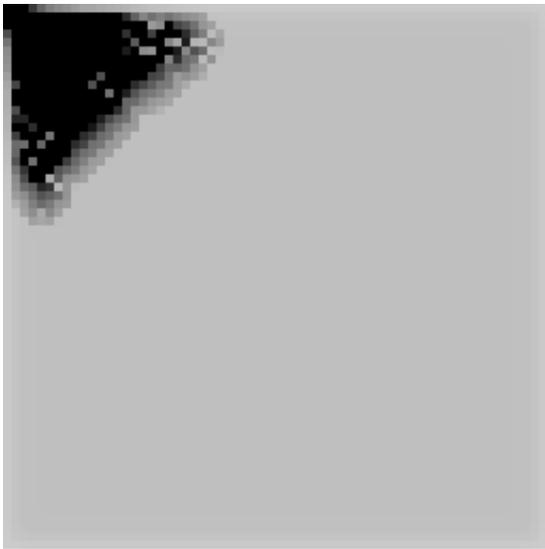
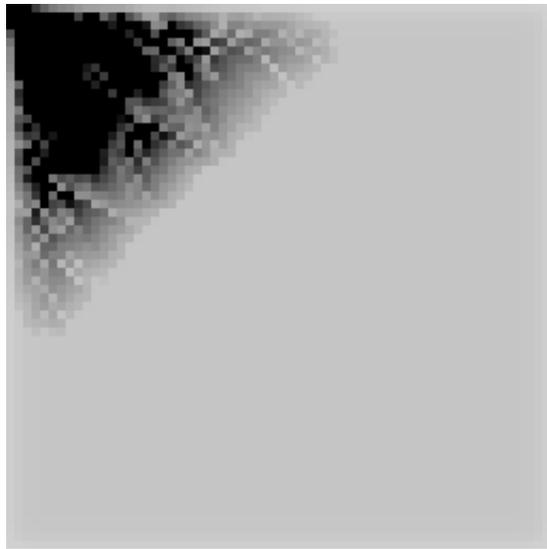
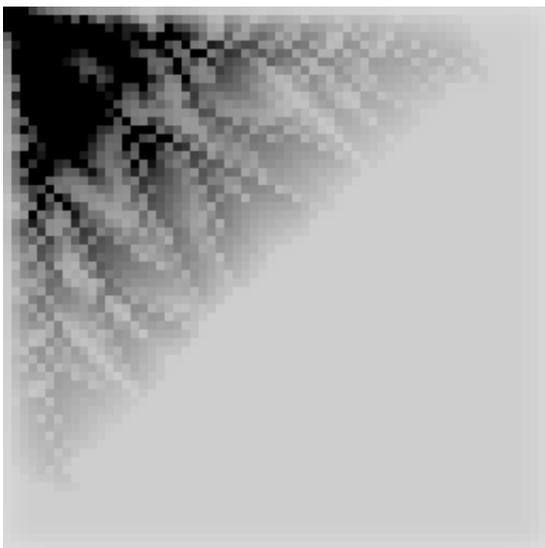
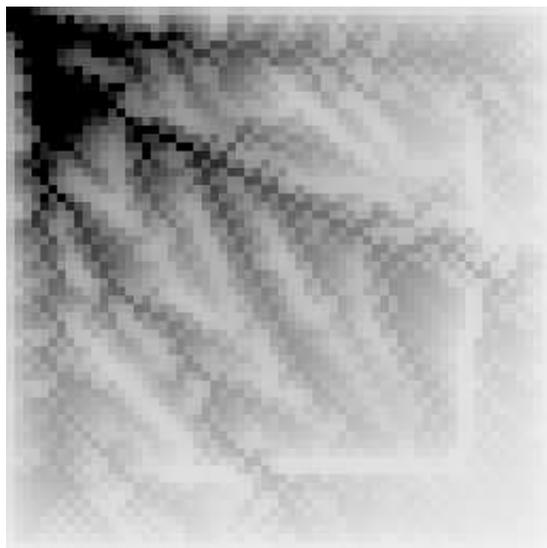
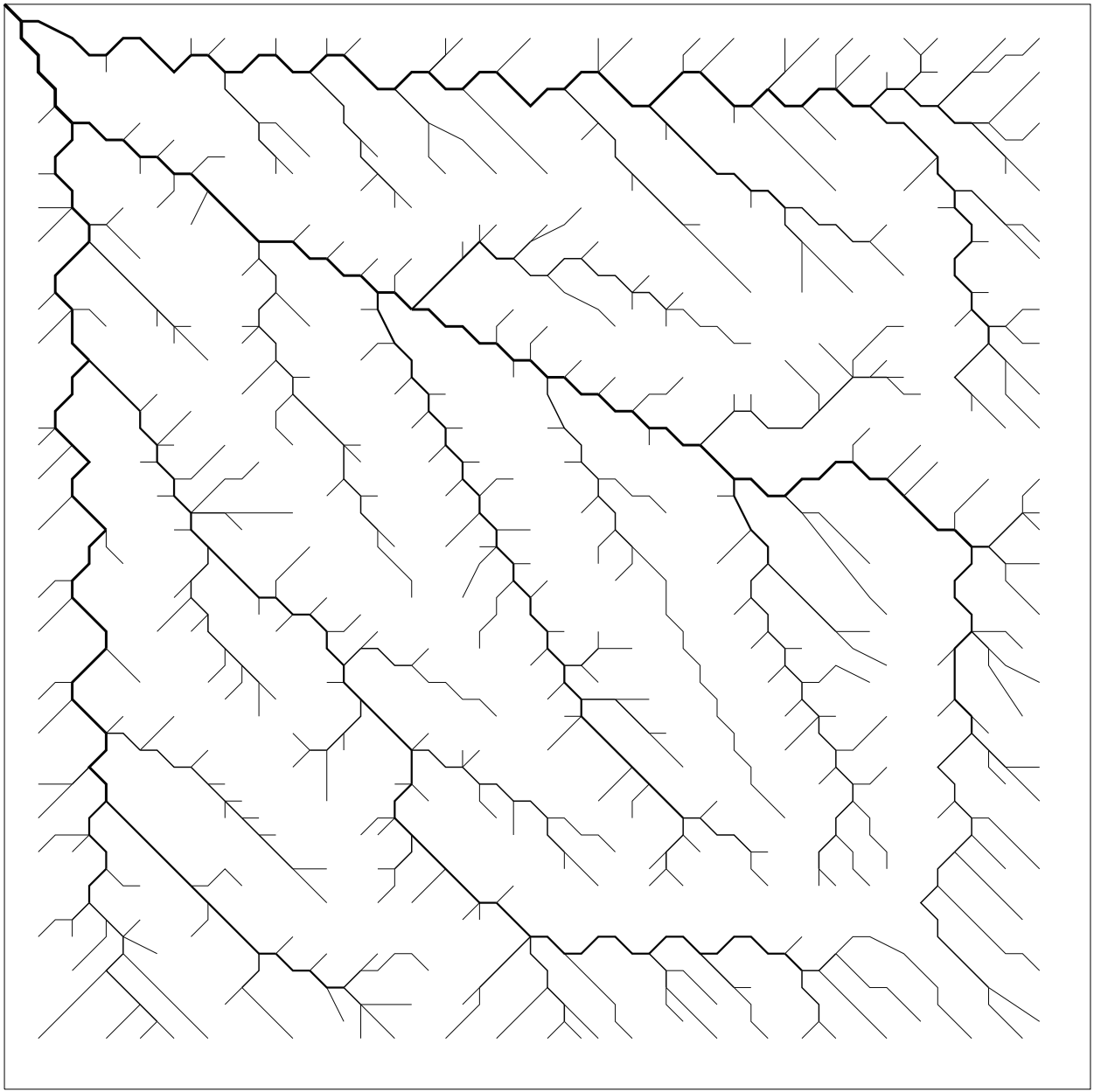
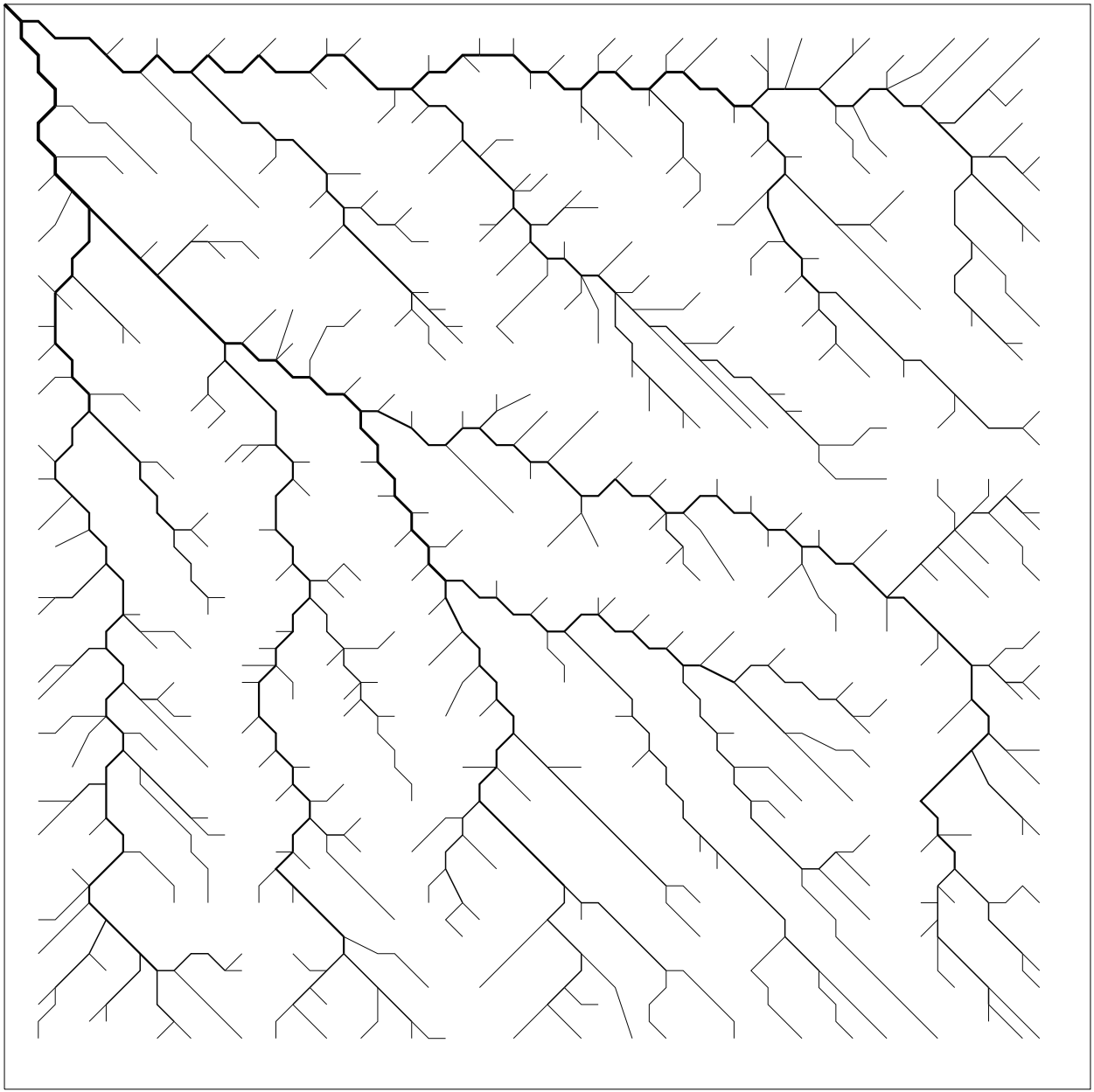
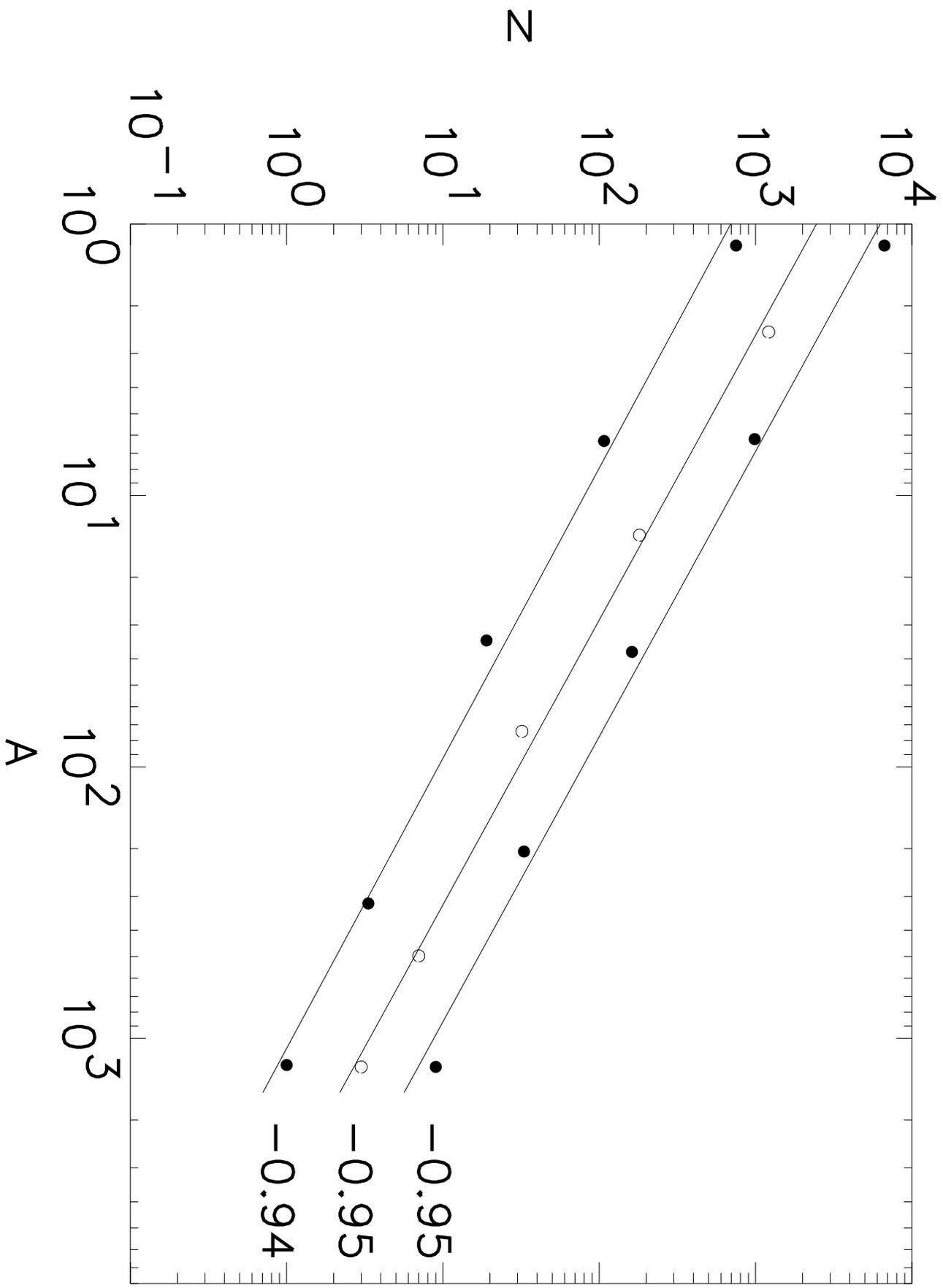
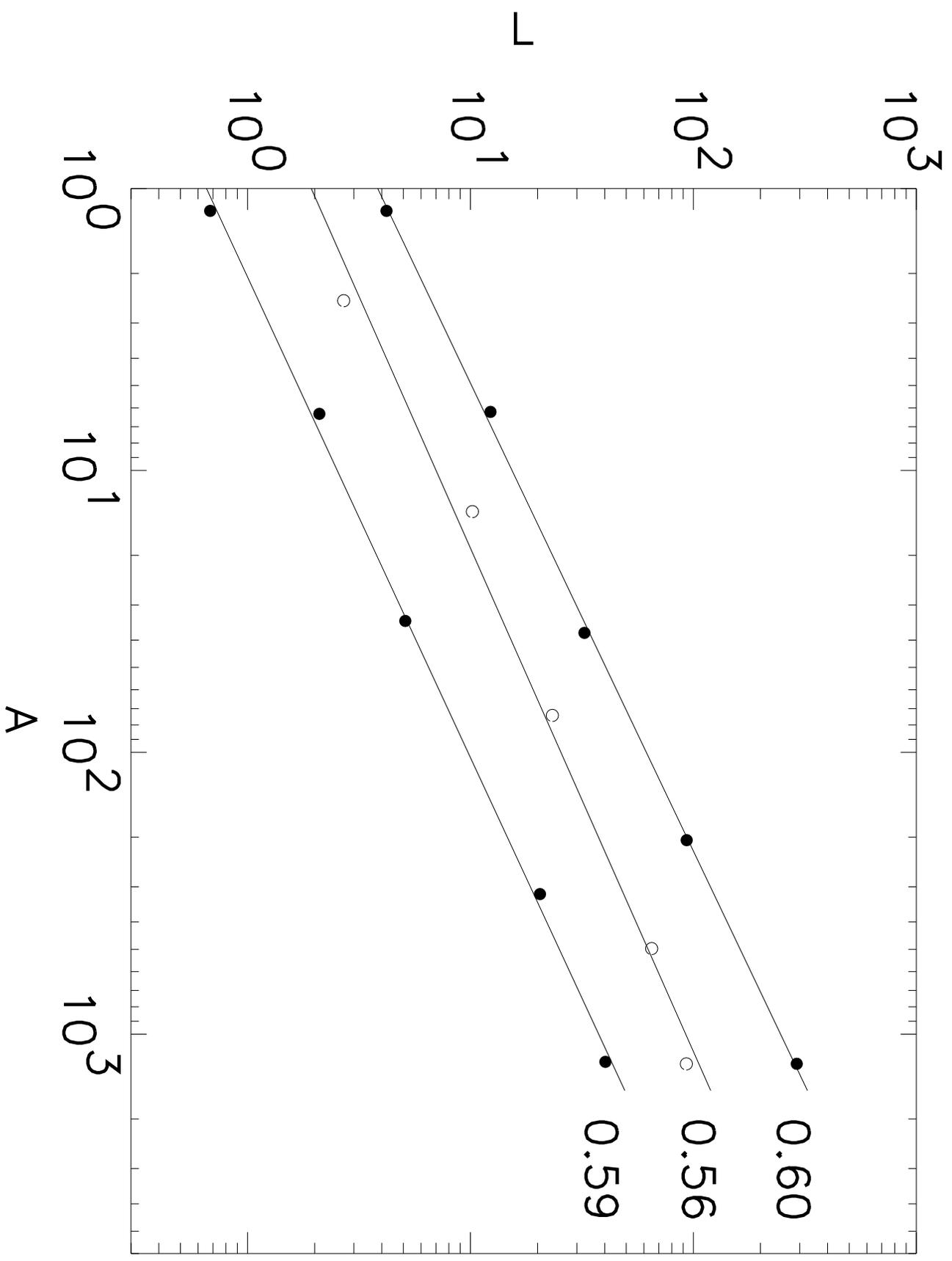
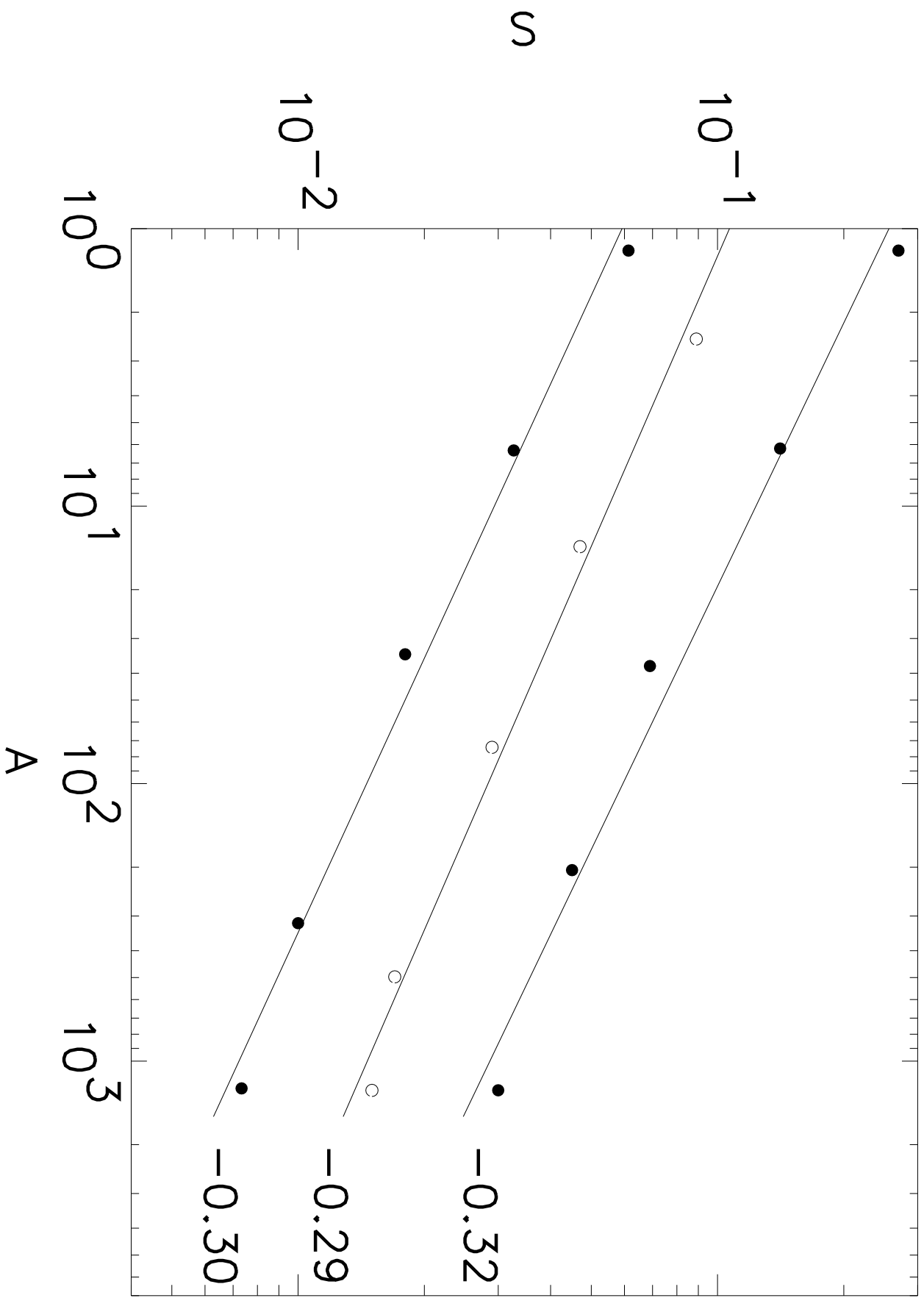
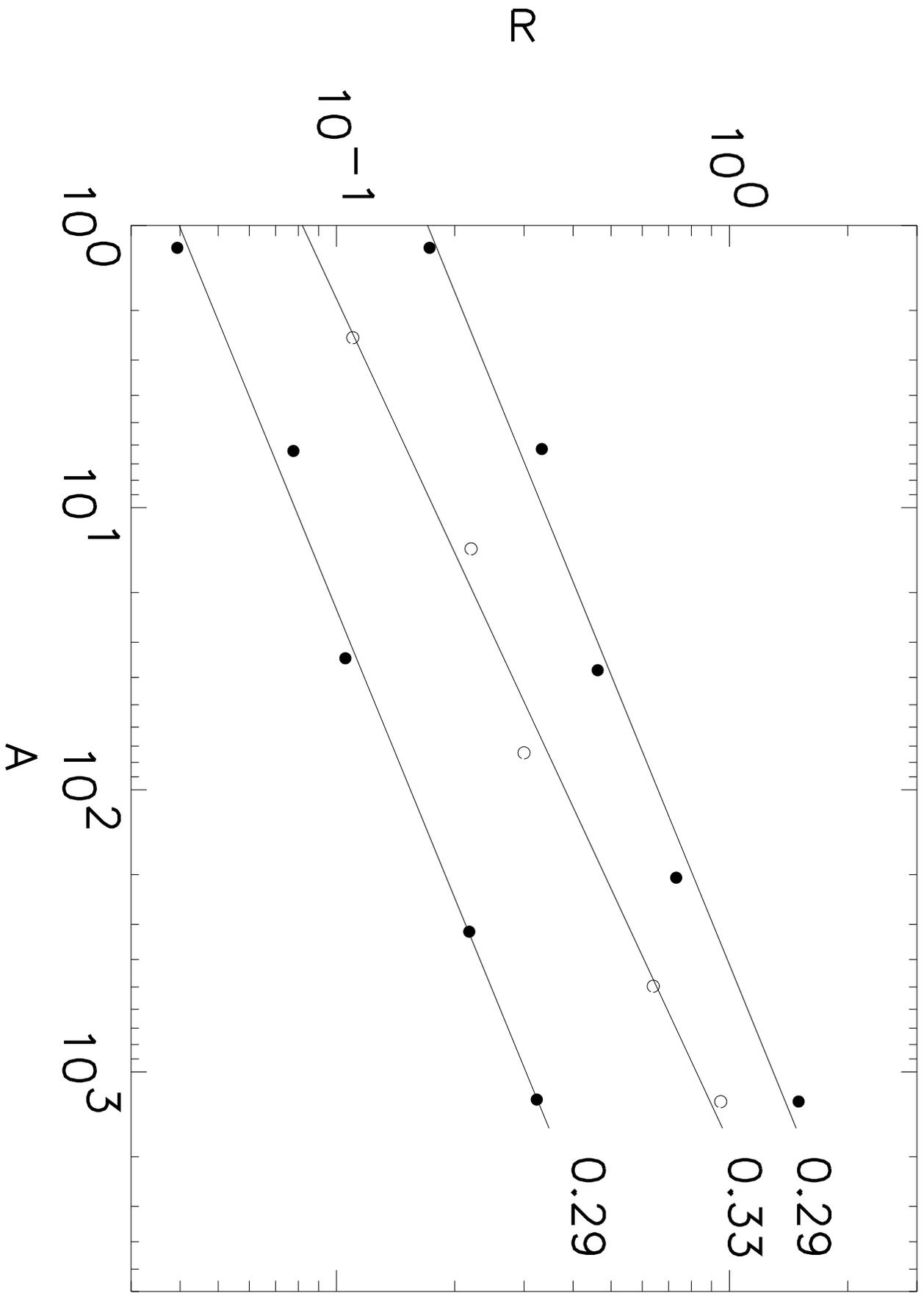
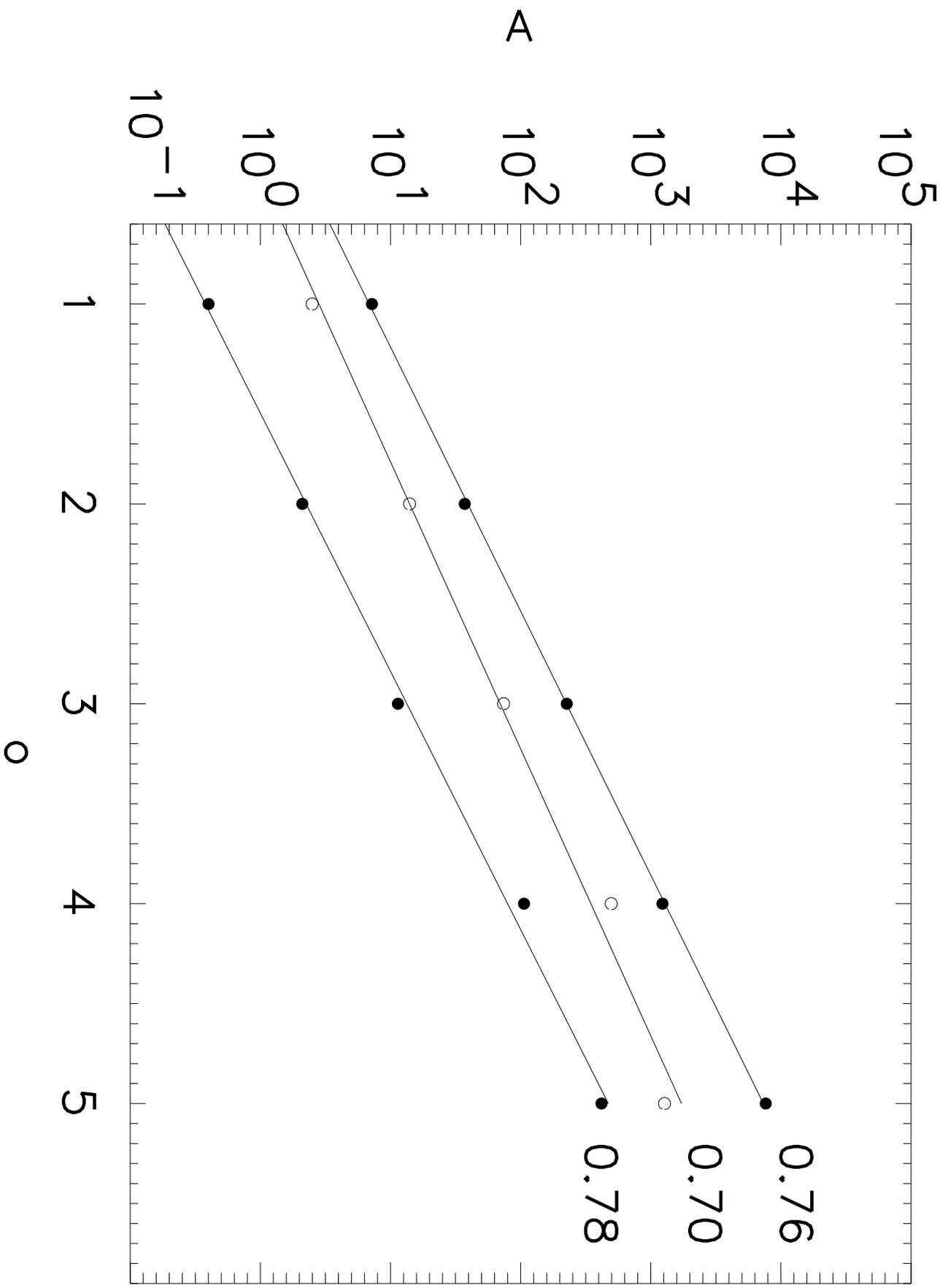
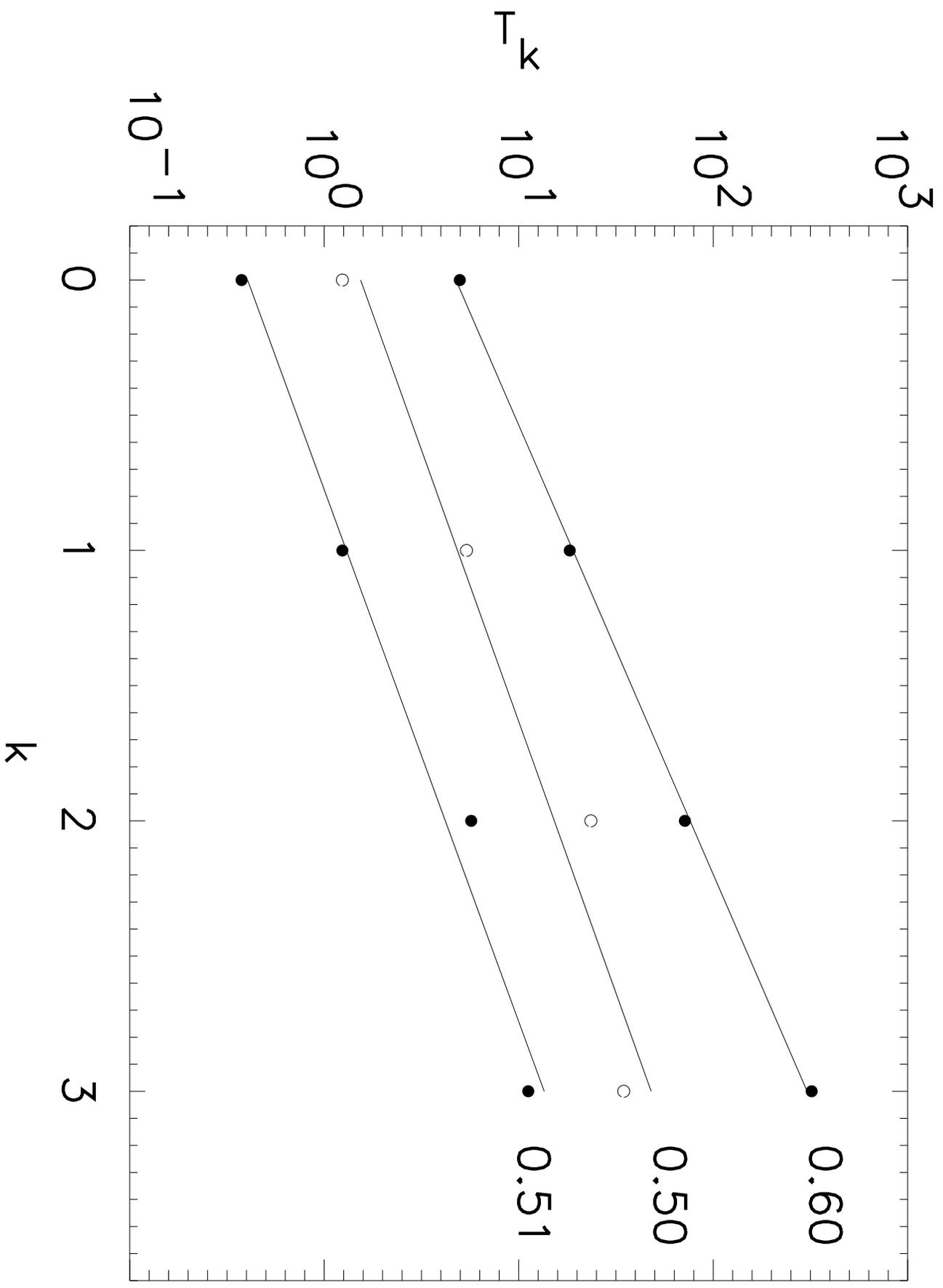
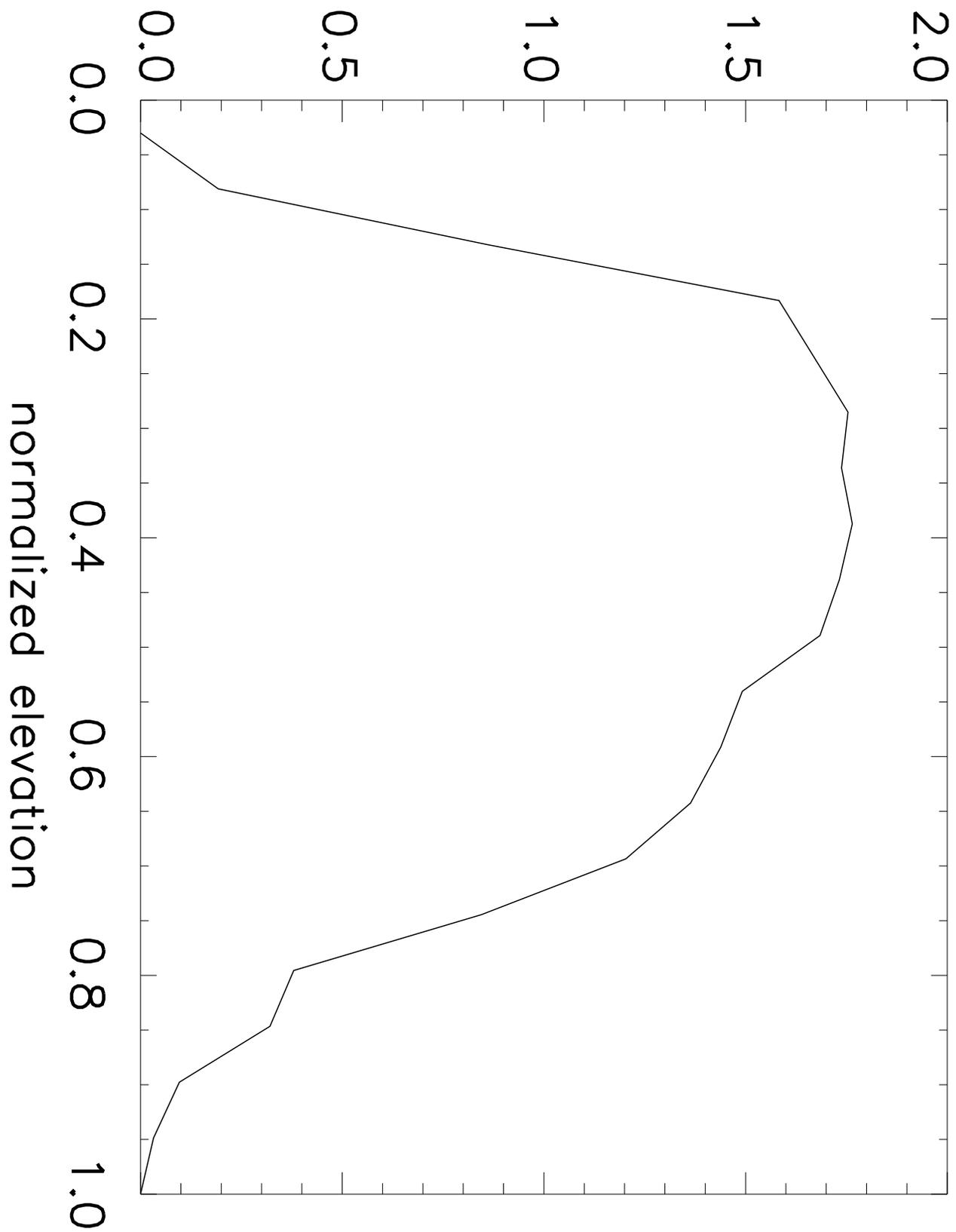
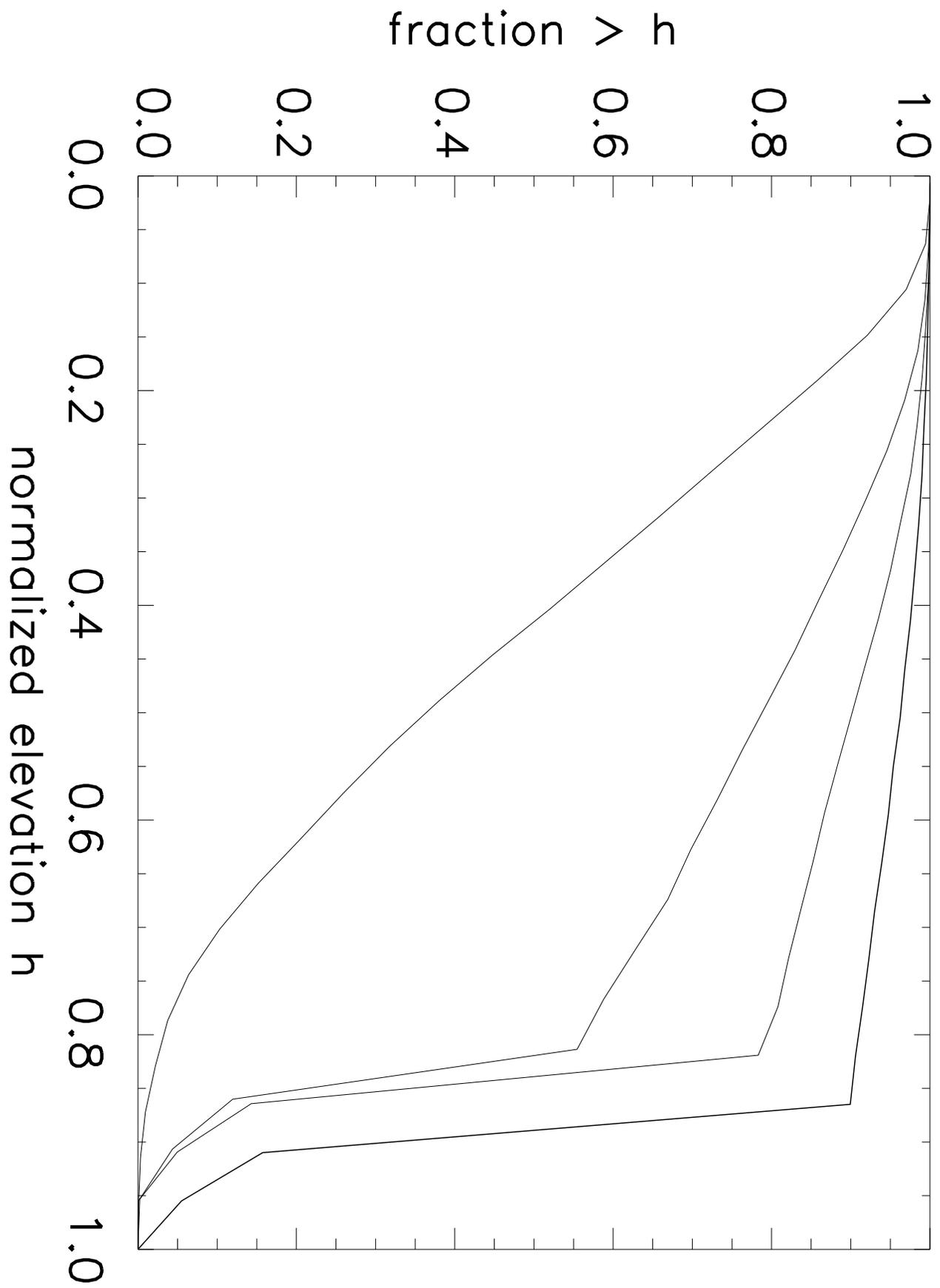
\end{document}